\documentclass[a4paper,fleqn]{cas-sc}

\usepackage[authoryear,longnamesfirst]{natbib}
\usepackage{amsmath,amssymb,amsfonts}
\usepackage{array}
\usepackage{algorithmic}
\usepackage{graphicx}
\usepackage{textcomp}
\usepackage{multirow}
\usepackage{wrapfig}
\usepackage{float}
\usepackage{tabularx}
\usepackage{pifont}
\usepackage{makecell}
\usepackage{rotating}
\usepackage{colortbl}
\usepackage{xcolor}
\usepackage{mathdots}
\usepackage{booktabs}
\usepackage{adjustbox}
\usepackage[inline]{enumitem}

\renewcommand{\arraystretch}{1.3}
\newcolumntype{R}{>{\raggedleft\arraybackslash}X}
\newcolumntype{Y}{>{\centering\arraybackslash}X}

\makeatletter
\def\maxwidth{\ifdim\Gin@nat@width>\linewidth\linewidth\else\Gin@nat@width\fi}
\def\maxheight{\ifdim\Gin@nat@height>\textheight\textheight\else\Gin@nat@height\fi}
\makeatother
\setkeys{Gin}{width=\maxwidth,height=\maxheight,keepaspectratio}

\begin{document}
	
\let\WriteBookmarks\relax
\def\floatpagepagefraction{1}
\def\textpagefraction{.001}

\shorttitle{MRUD Number Systems for High-Performance Arithmetic}
\shortauthors{S. Rahimi et al.}

\title[mode=title]{MRUD: Maximally Redundant Unsigned Digit Number Systems for High-Performance Arithmetic Units}

\author[1]{Samaneh Rahimi}[
orcid=0009-0000-6176-1645
]
\ead{rahimi@znu.ac.ir}
\credit{Conceptualization, Methodology, Software, Writing -- original draft}

\author[1]{Ali Azarpeyvand}[
orcid=0000-0002-4166-7528
]
\cormark[1]
\ead{azarpeyvand@znu.ac.ir}
\credit{Supervision, Writing -- review and editing}

\affiliation[1]{organization={Department of Electrical and Computer Engineering, University of Zanjan},
	city={Zanjan},
	postcode={38791-45371},
	country={Iran}}

\cortext[1]{Corresponding author}

\begin{abstract}
	Signed digit redundant number systems such as Maximally Redundant Signed Digit (MRSD) enable carry-free addition but introduce significant overhead in multiplication due to the presence of both positive and negative bits. While Unsigned Digit Redundant (UDR) systems have shown promise in radix-2, their extension to higher radices remains unexplored. This work presents the first generalized Maximally Redundant Unsigned Digit (MRUD) number systems for high-radix arithmetic, extending UDR representations from radix-2 to radix-$2^h$ while preserving carry-free addition. The proposed framework includes digit encoding, digit extension, arithmetic operations (addition, subtraction, and multiplication), sign inversion, forward/reverse conversion algorithms, overflow/underflow detection, and hardware architectures for both signed and unsigned MRUD representations. Unlike MRSD, which is inherently limited to signed arithmetic, MRUD supports both signed and unsigned representations, similar to conventional binary systems. Evaluations across three multiplier modes---RRR (Redundant inputs, Redundant output), BBR (Binary inputs, Redundant output), and BBB (Binary inputs, Binary output)---as well as Multiply-Accumulate (MAC) units, using the FreePDK 45nm library, demonstrate that MRUD achieves up to 30\% lower delay, 26\% smaller area, and 87\% lower power than MRSD. In BBR mode, MRUD achieves up to 43\% delay reduction over the best conventional architecture for 32$\times$32 operands, while in Multiply-Accumulate (MAC) units, performing reverse conversion only once at the end yields up to 64\% higher throughput than DesignWare and over 150\% higher than Wallace. These results demonstrate that MRUD enables high-radix redundant arithmetic with significantly improved multiplier and MAC performance, making it an efficient alternative to conventional signed digit redundant number systems.
\end{abstract}
\begin{highlights}
	\item First generalized unsigned digit redundant framework for high-radix arithmetic.
	\item Eliminating negabits simplifies partial product generation vs. MRSD.
	\item MRUD supports both signed and unsigned arithmetic, unlike MRSD.
	\item BBR-MRUD cuts delay up to 43\% versus best conventional multiplier.
	\item MRUD-based MAC achieves up to 64\% higher throughput than DesignWare.
\end{highlights}

\maketitle
	
	\section{Introduction}\label{introduction}
	
	Nowadays, with the growing integration of embedded systems into modern life, there is an increasing need for fundamental improvements in algorithms and architectures. This necessity arises from inherent limitations of conventional binary weighted number systems, which suffer from long carry propagation chains \citep{Molahisseini2017}. This issue becomes particularly critical in Very Large-Scale Integration (VLSI) environments---such as cryptography and Digital Signal Processing (DSP)---where arithmetic operations like addition and multiplication are key performance bottlenecks \citep{timarchi2022}. In binary systems, the delay of a Ripple Carry Adder (RCA) increases linearly with operand bit-width due to carry propagation \citep{timarchi2009}, and arithmetic performance deteriorates as operand bit-width increases \citep{parhami2009}. Consequently, developing alternative number systems has become an active research area aimed at improving speed and energy efficiency.
	
	Various alternative number systems have been explored, including the Logarithmic Number System (LNS) \citep{Parhami2020, Zhang2024}, the Residue Number System (RNS) \citep{Valueva2020}, and the Redundant Number System (RDNS) \citep{gorgin2016, Jaberipur2017, amanollahi-extended2018, Ghazanfari2021, amanollahi-AMR2022}. Each offers unique strengths and trade-offs: LNS simplifies multiplication and division but complicates addition and subtraction \citep{Molahisseini2017, Parhami2020, yu1991, huang1996, belanger2006}; RNS facilitates addition, subtraction, and multiplication but poses challenges in magnitude comparison and division \citep{timarchi2022}. Among these, RDNS has the unique ability to eliminate carry propagation in addition and subtraction under specific conditions, enabling carry-free addition with constant delay (\(O(1)\)) independent of operand bit-width \citep{Jaberipur2017}. However, redundant-digit multiplication remains \(O(\log n)\), offering no significant speed improvement over non-redundant counterparts \citep{amanollahi-extended2018}. Hybrid approaches such as the Redundant Residue Number System (RRNS) have also been explored to integrate the benefits of both RDNS and RNS \citep{Phalakarn2018, akbarzadeh2019, deng2024}.
	
	In RDNS, two signed digit representations are commonly used: Binary Signed Digit (BSD) and Maximally Redundant Signed Digit (MRSD) in radix-16. Both employ Signed Digits (SDs) based on the belief that Two's Complement (TC) encoding leads to more efficient adder designs \citep{timarchi2009, Jaberipur2017, Jaberi-ghodsi2003, Jaberi-gorginDesign2011, gorgin-afamily2011, moghaddasi-accelerator2024}. In terms of performance, BSD adders are faster than MRSD adders but occupy more area. In contrast, MRSD adders are slower than BSD adders but require less area. Additionally, storing data in BSD representation consumes more memory due to the extra bits. Consequently, in some computing systems, high-radix MRSD (e.g., radix-16) is considered an attractive option.
	
	In contrast to signed digit systems, Unsigned Digit Redundant (UDR) number systems were first introduced by \citet{parhami-ageneral1987} and further investigated in \citep{kawahito1991, parhami-comments1996, phatak-constant-timeAdd2001}. Specifically, \citet{phatak-constant-timeAdd2001} demonstrated that in radix-2, UDR adders---particularly CS2 (Carry-Save with digit set \([0, 2]\)) and CS3 (Carry-Save with digit set \([0, 3]\))---outperform SD adders with digit sets \([-1, 1]\) and \([-2, 1]\).
	
	The preference for SD over UDR can be attributed to the fact that SD, especially in higher radices, reduces the magnitude of digits by nearly half in operations such as multiplication \citep{amanollahi-extended2018, amanollahi-AMR2022} and division \citep{carter1990, amanollahi-devider2016, ebergen-2015}, leading to simpler hardware structures. However, in the radix-16 MRSD multiplier proposed in \citep{amanollahi-AMR2022}, although a lower delay is achieved compared to a conventional Binary Number System (BNS) multiplier, this delay reduction is not significant enough to justify the increased energy consumption and design area. The stages of Partial Product Generation (PPG) and Partial Product Reduction (PPR) in this multiplier are affected by the sign (polarity) of bits in signed digits. To ensure that the output of the PPR stage matches the format of the input operands, it must first be converted to BSD format and then transformed back to radix-16 MRSD format using 4-bit adders. This additional step results in increased multiplier depth, critical path delay, and design area.
	
	The application of UDR numbers in higher radices remains unexplored. This work introduces Maximally Redundant Unsigned Digit (MRUD) number systems and compares them with MRSD numbers. Unsigned digit representations simplify arithmetic operations by eliminating the need for negative bits, thereby reducing logic complexity. The main contributions of this work are summarized as follows:
	
	\begin{itemize}
		\item \textbf{Definition of MRUD Variants:}
		A maximally redundant digit set, \(D=[0,\beta]\), is introduced for UDR number systems, covering both even and odd values of \(\beta\). A similar classification was proposed in \citep{Jaberipur2017} for MRSD number systems, where symmetric and asymmetric digit sets \(D = [-\alpha, \beta]\) were defined.
		
		\item \textbf{Classification and Representation:}
		The MRUD number system is classified into unsigned and signed formats. For each format, polynomial representations, representational ranges, and bidirectional conversion methods between binary and MRUD are developed.
		
		\item \textbf{Arithmetic Operations:}
		Polynomial formulations for MRUD arithmetic operations are developed, including:
		\begin{itemize}
			\item Digit extension methods for signed and unsigned MRUD representations, enabling operand size alignment in multiplication and addition;
			\item A hardware-efficient sign inversion technique with closed-form expressions;
			\item A unified adder/subtractor architecture with comprehensive detection of real and apparent overflow/underflow conditions;
			\item Dedicated signed and unsigned multiplier architectures operating in RRR (Redundant inputs, Redundant output), BBR (Binary inputs, Redundant output), and BBB (Binary inputs, Binary output) modes.
		\end{itemize}
		
		\item \textbf{Multiply-Accumulate (MAC) Architecture:}
		A sequential MRUD-based MAC architecture is developed by integrating the proposed BBR-MRUD multiplier with redundant-domain accumulation, where reverse conversion is deferred until the final output. This architecture exploits carry-free accumulation throughout the MAC operation while requiring only a single reverse conversion.
	\end{itemize}
	
	All experimental results were obtained using Synopsys Design Compiler 2018, ModelSim SE 2020.4, and the FreePDK 45\,nm standard-cell library \citep{nangate45nm}. The abbreviations used throughout this paper are summarized in Table~\ref{tab:table_0}.
	
	\begin{table}[htbp]
		\caption{Description of symbols, abbreviations, and acronyms.}
		\label{tab:table_0}
		\centering
		\tiny
		\begin{tabular}{>{\centering\arraybackslash}m{3cm}>{\centering\arraybackslash}m{6cm}}
			\toprule
			\textbf{Symbols \& Abbreviations} & \textbf{Description} \\
			\midrule
			\ding{108} & Posibit (positive bit) \\
			\ding{109} & Negabit (negative bit) \\
			BBB & Binary inputs and binary output \\
			BBR & Binary inputs and redundant output \\
			BNS & Binary number system \\
			BSD & Binary signed digit \\
			CS2 & Carry-save with digit set \([0, 2]\) \\
			CS3 & Carry-save with digit set \([0, 3]\) \\
			DSP & Digital signal processing \\
			FA & Full adder \\
			GSD & Generalized signed digit \\
			LNS & Logarithmic number system \\
			MAC & Multiply-accumulate \\
			MRSD & Maximally redundant signed digit \\
			MRUD & Maximally redundant unsigned digit \\
			MSB & Most significant bit \\
			MSD & Most significant digit \\
			O & Algorithmic order \\
			PPG & Partial product generation \\
			PPR & Partial product reduction \\
			RCA & Ripple carry adder \\
			RDNS & Redundant number system \\
			RNS & Residue number system \\
			RRNS & Redundant residue number system \\
			RRR & Redundant inputs and redundant output \\
			SD & Signed digit \\
			SDF & Standard delay format \\
			TC & Two's complement \\
			UDR & Unsigned digit redundant \\
			VLSI & Very large-scale integration \\
			WBS & Weighted bit set \\
			\bottomrule
		\end{tabular}
	\end{table}
	
	The remainder of this work is organized as follows. Section~\ref{sec:Background} reviews background on signed digit redundant number systems. Section~\ref{sec:MRUD} introduces MRUD number systems and their mathematical foundations. Section~\ref{sec:Conversion} presents conversion methods between MRUD and binary representations. Section~\ref{sec:sign_inversion} derives the sign inversion method, while Section~\ref{sec:Unified} presents the unified addition/subtraction architecture. Section~\ref{sec:Multiplication} describes the proposed multiplication algorithms and hardware architectures. Section~\ref{sec:Experimental_Results} presents the experimental results, and Section~\ref{sec:Conclusion} concludes the work.
	
\section{Background}\label{sec:Background}

In radix-\(r\) conventional (non-redundant) number systems, each digit ranges from \(0\) to \(r-1\), forming a digit set of size \(r\): \([0, r-1]\). When the cardinality of a digit set exceeds the radix \(r\), the number system is classified as a RDNS. These redundant digits can be positive or negative, allowing multiple representations of certain numbers. Consequently, addition operations can be performed as carry-limited or carry-free \citep{Jaberipur2017}.

RDNS with symmetric signed digit sets, defined as \(D = [-\alpha, \alpha]\) with \(\alpha \geq \lceil (r+1)/2 \rceil\), were first introduced by Avizienis \citep{Avizienis1961}. Three decades later, Parhami extended this concept to Generalized Signed Digit (GSD) number systems by defining a broader digit set, \(D = [-\alpha, \beta]\) \citep{parhami-generalized1990}. He also introduced the redundancy index \(\rho=\xi_{\text{D}}-r\), where \(\xi_{\text{D}}\) denotes the cardinality of the digit set. In GSD, carry-free addition is not always guaranteed. According to \citet{Jaberipur2017}, for a digit set \(D = [-\alpha, \beta]\), carry-free addition is possible if either:
\begin{enumerate}
	\item \(r > 2\) and \(\rho \geq 3\), or
	\item \(r > 2\), \(\rho = 2\), with \(\alpha \neq 1\) and \(\beta \neq 1\).
\end{enumerate}
GSD encompasses all cases of redundant digit sets, including those with maximum redundancy (\(D_{\text{max}}\)). Maximum redundancy indicates that \(D_{\text{max}}\) has the maximum cardinality while the minimum number of bits are used to represent its digits.

\begin{figure}[pos=htbp]
	\centering
	\includegraphics[width=0.3\columnwidth, keepaspectratio=true]{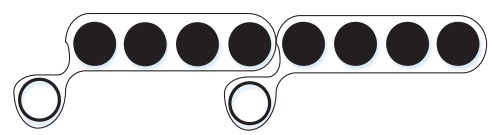}
	\caption{Two consecutive radix-16 digits in the MRSD number system \citep{amanollahi-extended2018}.}
	\label{fig_mrsd_digits}
\end{figure}

In general, there are two types of radix-\(2^h\) MRSD number systems: (1) MRSD with a symmetric digit set \(D=[-2^h+1, 2^h-1]\) and (2) MRSD with an asymmetric digit set \(D=[-2^h, 2^h-1]\). Fig.~\ref{fig_mrsd_digits} illustrates two consecutive radix-16 digits in the MRSD number system. In this figure, positive bits (posibits) are represented by the symbol \ding{108}, and negative bits (negabits) by the symbol \ding{109}. The logical values of 0 and 1 for posibits correspond directly to their arithmetic values, whereas for negabits, a logical 0 represents an arithmetic value of \(-1\), and a logical 1 represents an arithmetic value of 0.

The MRSD number system offers a trade-off between delay and area consumption, enabling optimization based on application requirements. As the parameter \(h\) in MRSD addition decreases, delay reduces at the cost of increased area, and conversely, increasing \(h\) lowers area consumption but extends delay \citep{gorgin-afamily2011}. This flexibility makes MRSD suitable for diverse embedded computing applications.

In \citet{Jaberipur2017}, encoding efficiency is defined as \(\epsilon=\xi_{\text{D}}/2^b\), where \( b \) represents the number of bits required to encode these digits. This parameter highlights that the MRSD with an asymmetric digit set offers higher encoding efficiency than MRSD with a symmetric digit set.

The MRSD with symmetric digit set employs an unfaithful digit set encoding, enabling it to represent \((-2^h)\)---a value absent from its digit set---resulting in additional cost and delay \citep{amanollahi-extended2018}. Consequently, MRSD with an asymmetric digit set is the preferred choice for designing computational circuits based on MRSD number systems, as demonstrated in \citep{amanollahi-extended2018, amanollahi-AMR2022}.

While MRSD offers flexibility in the delay-area trade-off, the presence of signed digits introduces complexity in arithmetic operations, particularly in multiplication. This motivates the exploration of unsigned digit redundant representations, which are introduced as MRUD number systems in the following section.

\section{MRUD Number Systems}\label{sec:MRUD}

In general, arithmetic operations with posibits are simpler and more hardware-efficient than operations involving a mix of posibits and negabits. A representative example is the 4:2 compressor, which uses only posibits. This compressor can be viewed as a redundant unsigned digit adder where each operand has a 2-deep representation with \(r=2\) and \(D=[0,2]\). Consequently, carry-free addition is not possible, and carry-limited addition is employed instead.

\begin{figure}[pos=htbp]
	\centering
	\scriptsize
	\begin{tabular}{cc}
		\includegraphics[width=0.3\textwidth, keepaspectratio=true]{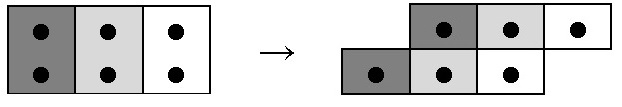}&\includegraphics[width=0.3\textwidth, keepaspectratio=true]{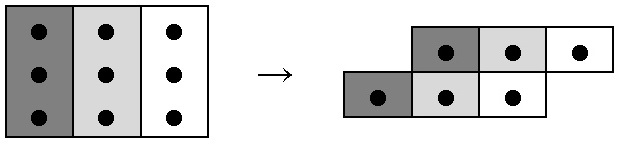}\\
		(a) \(D=[0,2], r=2, \rho=1\) (minimally redundant radix-2)&(b) \(D=[0,3], r=2, \rho=3\) (maximally redundant radix-2)\\
		\includegraphics[width=0.3\textwidth, keepaspectratio=true]{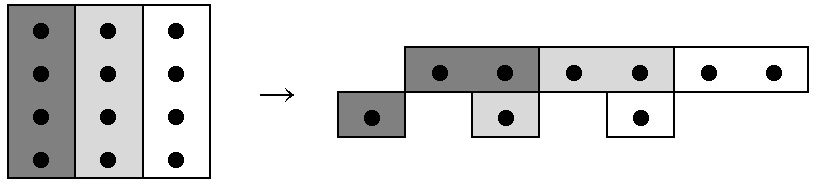}&\includegraphics[width=0.3\textwidth, keepaspectratio=true]{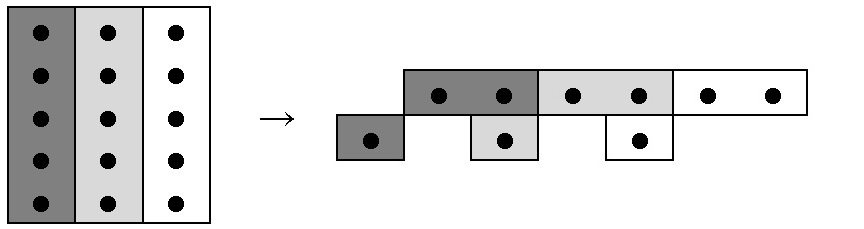}\\
		(c) \(D=[0,4], r=4, \rho=1\) (minimally redundant radix-4)&(d) \(D=[0,5], r=4, \rho=2\)\\
		\includegraphics[width=0.3\textwidth, keepaspectratio=true]{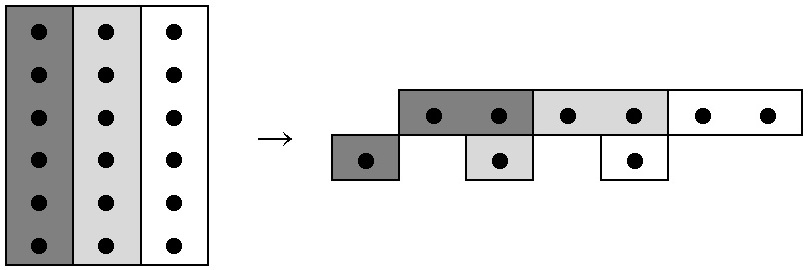}&\includegraphics[width=0.3\textwidth, keepaspectratio=true]{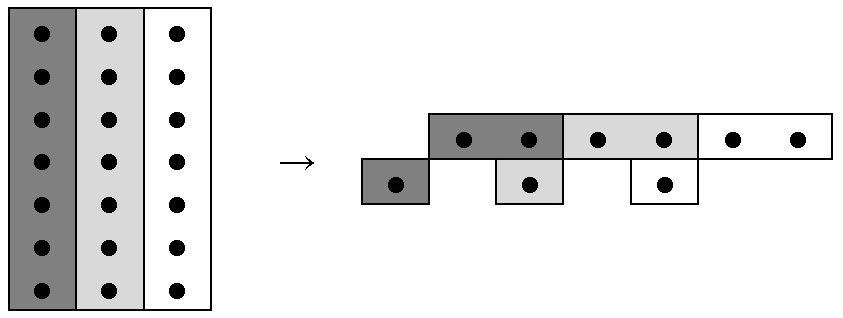}\\
		(e) \(D=[0,6], r=4, \rho=3\)&(f) \(D=[0,7], r=4, \rho=4\) (maximally redundant radix-4)
	\end{tabular}
	\caption{2-deep WBS encoding of a 3-digit UDR operand for different digit sets: (a) minimally redundant radix-2, (b) maximally redundant radix-2, (c) minimally redundant radix-4, (d)--(e) intermediate redundancy radix-4, (f) maximally redundant radix-4.}
	\label{fig-udr}
\end{figure}

Fig.~\ref{fig-udr}(a)--(f) illustrates how operands with various digit sets are converted into 2-deep operands using the Weighted Bit Set (WBS) encoding proposed in \citep{JaberiWBS2002}. Fig.~\ref{fig-udr}(a) and (b) correspond to radix-2, while Fig.~\ref{fig-udr}(c)--(f) correspond to radix-4. As observed in the figure, carry-free addition is possible only for radix-4 digit sets with sufficient redundancy, i.e., Fig.~\ref{fig-udr}(d)--(f), where \(r=4\) and \(\rho \geq 2\).

After constructing the equivalent higher-radix 2-deep representation for operands in multi-operand adders, it is essential to define an appropriate digit set for the UDR number system. According to \citet{parhami-on-the-impl1993}, the complexity increase in a GSD system is negligible when \(\alpha \equiv \beta \pmod{r-1}\). In UDR, \(\alpha = 0\), so \(\beta = q(r-1)\) for some integer \(q\). For \(q = 1\), \(\beta = r-1\), which is non-redundant. For \(q \geq 2\), the system becomes redundant. Choosing the minimal value \(q = 2\) ensures negligible complexity increase, yielding the digit set \(D = [0, 2(r-1)]\).

Similar to the SD number system, the digit set of the UDR number system is defined for both minimum and maximum redundancy as follows:
\begin{itemize}
	\item \textbf{Definition 1 -- Minimum Redundancy (\(D_{\text{min}}\))}:
	The digit set \(D_{\text{min}}\) exhibits minimum redundancy when \(\rho_{\text{min}} = 1\). The binary representation and 2-deep encoding of such digits require at least \(b_{\text{min}} = \lceil \log_2 \xi_{D_{\text{min}}} \rceil\) bits. For \(r = 2^h\), this simplifies to \(b_{\text{min}} = h + 1\). Fig.~\ref{fig-udr}(a) and (c) illustrate minimally redundant digit sets for radices 2 and 4, respectively.
	
	\item \textbf{Definition 2 -- Maximum Redundancy (\(D_{\text{max}}\))}:
	The digit set \(D_{\text{max}}\) achieves maximum redundancy if:
	\[
	\xi_{D_{\text{max}}} =
	\begin{cases}
		2^{b_{\text{min}}}, & \text{if } \beta \text{ is odd},\\
		2^{b_{\text{min}}} - 1, & \text{if } \beta \text{ is even}.
	\end{cases}
	\]
	This ensures that \(D_{\text{max}}\) contains the maximum possible number of elements while still using the minimum number of bits for the 2-deep representation of its digits, as shown in Fig.~\ref{fig-udr}(b) and (f).
\end{itemize}

According to Definition 2, when the digit set is unfaithful with \(\beta = 2^{h+1} - 2\) (which is even), i.e., \(D = [0, 2^{h+1} - 2]\), this digit set achieves maximum redundancy. For example, if \(h = 4\) (radix-16), using four bits instead of five to represent the digits makes the digit set non-redundant. On the other hand, for \(r = 2^h\), the digit set \(D = [0, 2^{h+1} - 2]\) takes the form \(D = [0, 2(r-1)]\), resulting in minimal complexity increase.

The most important feature of MRUD number systems is that they only deal with positive bits and do not use negative bits. This not only simplifies working with the bits but also suggests that many arithmetic operations commonly implemented in MRSD can be adapted to MRUD with minor modifications. Furthermore, like conventional number systems, MRUD can be classified into signed and unsigned forms, enabling application-specific optimizations.

The following sections present the conversion methods and arithmetic operations for MRUD number systems.

\subsection{Signed MRUD Number System}\label{sec:smrud}

The polynomial representation for a signed MRUD number system with radix \(r\) and digit set \(D = [0, \beta]\), denoted as \((x)_{r, \beta}\), is given in \eqref{equ1}. In this representation, the Most Significant Digit (MSD) has an effective weight of \((x_{n-1} - r)\), where the negative component \(-r\) enables the representation of negative numbers, while the remaining digits have conventional positive weights \(r^i\):

\begin{equation}
	\label{equ1}
	(X)_{10} = (x)_{r, \beta} = (x_{n-1} - r) r^{n-1} + \sum_{i=0}^{n-2} x_i r^i, 
	\qquad 0 \leq x_i \leq \beta
\end{equation}

where \(\beta = r-1 + \rho\), \(n\) is the number of digits, and \(r\) is the radix. The representable range of signed MRUD numbers is derived from \eqref{equ1} as:

\begin{equation}
	\label{equ2}
	-r^n \leq X \leq -r^n + \frac{\beta}{r-1}(r^n - 1)
\end{equation}

Compared to the MRSD system, signed MRUD offers a wider range on the positive side (positive bias), whereas MRSD provides a wider range on the negative side. This positive bias in MRUD is a direct consequence of using only posibits and the negative weight assigned to the MSD.

\subsection{Unsigned MRUD Number System}\label{sec:umrud}

Many applications, such as digital image processing, involve only positive or unsigned numbers. The key advantage of using the unsigned MRUD number system is its ability to provide a wider representable range for positive numbers, enabling fewer digits and consequently reducing word length and area.

The polynomial representation for an unsigned MRUD number system with radix \(r\) and digit set \(D = [0, \beta]\), denoted as \((x)_{r, \beta}\), is given in \eqref{equ3}, and its representable range is given in \eqref{equ4}:

\begin{equation}
	\label{equ3}
	(X)_{10} = (x)_{r, \beta} = \sum_{i=0}^{n-1} x_i r^i, \qquad 0 \leq x_i \leq \beta
\end{equation}

\begin{equation}
	\label{equ4}
	0 \leq X \leq \frac{\beta}{r-1}(r^n - 1)
\end{equation}
The polynomial relation for the unsigned MRUD number system (i.e., \eqref{equ3}) is similar to that of conventional unsigned numbers, with the difference that the digit range is \(0 \leq x_i \leq \beta\) instead of \(0 \leq x_i \leq r-1\).

A comparison of the representable ranges for the signed and unsigned MRUD number systems, as derived in \eqref{equ2} and \eqref{equ4}, reveals two key findings:
\begin{enumerate}
	\item When \(\beta\) is odd, the representable range is wider than when \(\beta\) is even.
	\item For the signed MRUD system with even \(\beta\), the range exhibits greater symmetry.
\end{enumerate}
However, an odd \(\beta\) is preferred for both the signed and unsigned MRUD systems due to the increased number of representable values, enhancing their versatility in arithmetic applications.

\subsection{Digit Extension in MRUD Number Systems}
\label{sec:digit_extension}

In arithmetic operations such as multiplication, the number of digits in the operands may differ from the number of digits required for the result. Consequently, it is often necessary to extend an $n$-digit MRUD number to $n+k$ digits without altering its arithmetic value. This property, known as digit extension, is fundamental to maintaining consistency across different operand sizes and ensuring compatibility between various arithmetic units.

\subsubsection{Unsigned MRUD Extension}

An $n$-digit unsigned MRUD number $(X)_{r,\beta} = \sum_{i=0}^{n-1} x_i r^i$ is extended to $n+k$ digits by appending $k$ zeros:
\begin{equation}
	\label{equ5}
	(X)_{r,\beta} = \sum_{i=0}^{n+k-1} x'_i r^i, \qquad
	x'_i =
	\begin{cases}
		x_i, & 0 \le i < n \\[4pt]
		0, & n \le i < n+k
	\end{cases}
\end{equation}
This follows directly from \eqref{equ3}, since the additional digits have zero weight. For instance, $(A_1A_0)_{r,\beta}$ extends to $(00A_1A_0)_{r,\beta}$.

\subsubsection{Signed MRUD Extension}

An $n$-digit signed MRUD number $(X)_{r,\beta} = -r^n + \sum_{i=0}^{n-1} x_i r^i$ is extended to $n+k$ digits by appending $k$ digits equal to $(r-1)$ as the new MSDs:
\begin{equation}
	\label{equ6}
	(X)_{r,\beta} = -r^{n+k} + \sum_{i=0}^{n+k-1} x'_i r^i, \qquad
    x'_i =
	\begin{cases}
		x_i, & 0 \le i < n \\
		r-1, & n \le i < n+k
	\end{cases}
	\end{equation}
The validity of this extension is confirmed by the following identity:
\[
\begin{split}
	(X)_{r,\beta} &= -r^n + \sum_{i=0}^{n-1} x_i r^i
	= -r^{n+k} + r^{n+k} - r^n + \sum_{i=0}^{n-1} x_i r^i 
	= -r^{n+k} + (r-1) \frac{r^{n+k} - r^n}{r-1} + \sum_{i=0}^{n-1} x_i r^i \\[4pt]
	&= -r^{n+k} + (r-1) \sum_{i=n}^{n+k-1} r^i + \sum_{i=0}^{n-1} x_i r^i
\end{split}
\]
For example, $(56)_{4,7}=10$ extends to $(3356)_{4,7}=10$.

\section{Conversion Between Conventional and MRUD Number Systems}\label{sec:Conversion}

\begin{figure}[pos=htbp]
	\centering
	\scriptsize
	\begin{tabular}{c}
		\includegraphics[width=0.7\columnwidth]{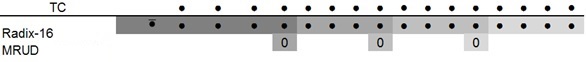}\\
		(a) From TC to signed MRUD.\\
		\\
		\includegraphics[width=0.7\columnwidth]{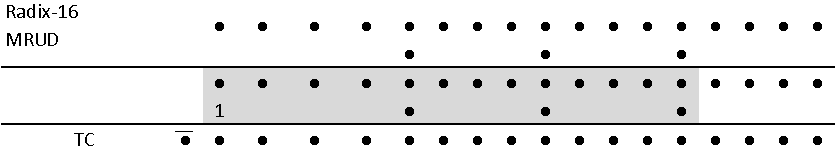}\\
		(b) From signed MRUD to TC.
	\end{tabular}
	\caption{Conversion operation between two's complement and signed MRUD.}
	\label{fig-conversion}
\end{figure}

Fig.~\ref{fig-conversion} illustrates the conversion processes between a 16-bit two's complement (TC) number and a 4-digit signed MRUD number in radix 16, as well as the reverse conversion.

\subsection{Direct Conversion (TC to Signed MRUD)}

As shown in Fig.~\ref{fig-conversion}(a), in the direct conversion from a 16-bit TC number to a 4-digit signed MRUD number, three posibits, each with a value of 0, are placed in positions with weights \(2^{4k}\) (where \(k \in \{1,2,3\}\)). The most significant bit (MSB) of the source TC number is positioned at weight \(2^{15}\), while its bitwise inverse is placed at weight \(2^{16}\). This direct conversion requires only a single inverter, making it hardware-efficient.

\subsection{Reverse Conversion (Signed MRUD to TC)}

For the reverse conversion from a 4-digit signed MRUD number to an 18-bit TC number, as depicted in Fig.~\ref{fig-conversion}(b), a bit with a value of 1 is first placed at position \(2^{16}\). The shaded bits in the figure are then fed into a 13-bit non-redundant adder. The carry-out from this 13-bit adder is inverted to serve as the most significant bit of the resulting 18-bit TC number. The output is an 18-bit TC number because the signed MRUD representation has a wider representable range than the original 16-bit TC format.

Since the reverse conversion involves carry propagation, it is advantageous to perform as many arithmetic operations as possible in the redundant domain before the final reverse conversion, thereby fully leveraging the carry-free nature of RDNS.

The direct and reverse conversion methods described here closely resemble those for MRSD \citep{Jaberipur2017}, with the key difference that MRUD uses only posibits, eliminating the need for negabit handling during conversion.

\subsection{Unsigned MRUD Conversion}

The conversion process for the unsigned MRUD number system closely resembles that of the signed MRUD system. In the direct conversion from an unsigned binary number to an unsigned MRUD number, bits with a value of 0 are placed in all positions with weights \(2^{4k'}\) (where \(k' \in \{1,2,3,4\}\)), incurring no additional cost. In the reverse conversion, a bit with a value of 0 is placed at position \(2^{16}\), followed by the use of the 13-bit adder. The carry-out from this 13-bit adder determines the last bit of the resulting unsigned binary number.

\section{Sign Inversion in MRUD Number Systems}\label{sec:sign_inversion}

From the polynomial representation of a signed MRUD number in \eqref{equ1}:
\begin{equation}
	\label{equ7}
	-(X)_{r,\beta} = r^n - \sum_{i=0}^{n-1} x_i r^i
\end{equation}

Let \(\overline{x_i} = \beta - x_i\) denote the digitwise complement. Substituting \(x_i = \beta - \overline{x_i}\) into \eqref{equ7} gives:

\begin{equation}
	\label{equ8}
	\begin{split}
		-(X)_{r,\beta} &= r^n - \sum_{i=0}^{n-1} (\beta - \overline{x_i}) r^i 
		= r^n - \beta \sum_{i=0}^{n-1} r^i + \sum_{i=0}^{n-1} \overline{x_i} r^i 
		= -r^n + 2r^n - \beta \cdot \frac{r^n - 1}{r-1} + \sum_{i=0}^{n-1} \overline{x_i} r^i \\
		&= -r^n + r^n \left(2 - \frac{\beta}{r-1}\right) + \frac{\beta}{r-1} + \sum_{i=0}^{n-1} \overline{x_i} r^i 
		= -r^n - r^n \left(\frac{\beta}{r-1} - 2\right) + \left(\frac{\beta}{r-1} - 2\right) + 2 + \sum_{i=0}^{n-1} \overline{x_i} r^i \\
		&= -r^n + \left(\frac{\beta}{r-1} - 2\right)(1 - r^n) + 2 + \sum_{i=0}^{n-1} \overline{x_i} r^i
	\end{split}
\end{equation}

For MRUD with \(\beta = 2(r-1)\) (even) and \(\beta = 2(r-1) + 1\) (odd), \eqref{equ8} simplifies to:

\begin{enumerate}
	\item For \(\beta = 2(r-1) + 1\):
	\begin{equation}
		\label{equ9}
		-(X)_{r,\beta} = -r^n + (r-2) \sum_{i=0}^{n-1} r^i + \sum_{i=0}^{n-1} \overline{x_i} r^i + 3
	\end{equation}
	
	\item For \(\beta = 2(r-1)\):
	\begin{equation}
		\label{equ10}
		-(X)_{r,\beta} = -r^n + \sum_{i=0}^{n-1} \overline{x_i} r^i + 2
	\end{equation}
\end{enumerate}

\begin{figure}[pos=htbp]
	\centering
	\includegraphics[width=0.5\columnwidth, keepaspectratio=true]{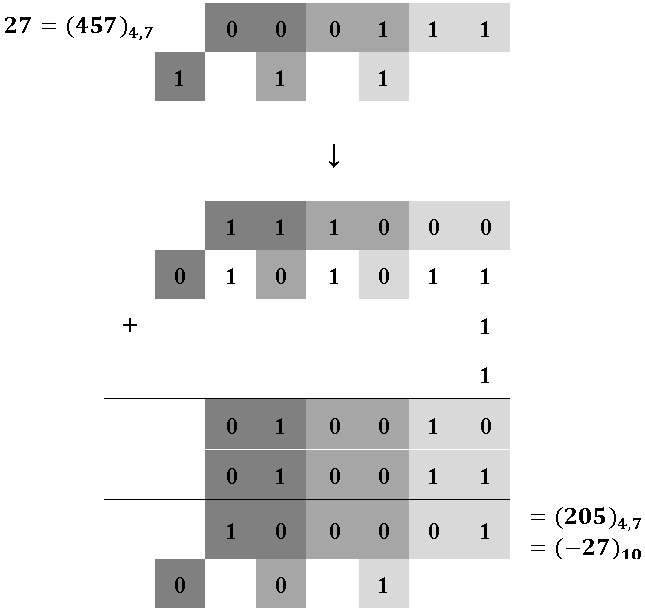}
	\caption{Sign inversion example for radix-4 MRUD with \(n=3\) and \(\beta=7\). The number \((457)_{4,7}\) equals \(27_{10}\), and its inversion \((205)_{4,7}\) equals \(-27_{10}\).}
	\label{fig_sign_inversion}
\end{figure}

Fig.~\ref{fig_sign_inversion} illustrates an example for \(r=4\), \(\beta=7\), and \(n=3\). The number \((457)_{4,7}\) is interpreted as:
\[
(457)_{4,7} = (4-4) \times 4^2 + 5 \times 4^1 + 7 \times 4^0 = 20 + 7 = 27_{10}.
\]

To compute its sign inversion, each digit is first complemented: \((457)_{4,7} \rightarrow (320)_{4,7}\). Then, as derived in \eqref{equ9}, a constant of \(r-2 = 2\) is added to each digit, followed by an overall constant of \(3\). This yields the sign-inverted value \(-27_{10}\), which in MRUD representation is \((205)_{4,7}\):
\[
(205)_{4,7} = (2-4) \times 4^2 + 0 \times 4^1 + 5 \times 4^0 = -32 + 5 = -27_{10}.
\]

\section{MRUD Addition and Subtraction Operations}\label{sec:Unified}

Similar to conventional number systems, addition and subtraction in MRUD may result in overflow/underflow. To analyze these conditions, two $n$-digit signed MRUD operands are considered:
\begin{equation}
	\label{equ11}
	X = -r^n + \sum_{i=0}^{n-1} x_i r^i, \quad 0 \le x_i \le \beta
\end{equation}
\begin{equation}
	\label{equ12}
	Y = -r^n + \sum_{i=0}^{n-1} y_i r^i, \quad 0 \le y_i \le \beta
\end{equation}

The result is assumed to be $n+1$ digits, and the conditions for overflow/underflow are subsequently examined.

\subsection{Addition Operation}\label{sec:add_op}

The sum of two signed MRUD numbers is given by:
\begin{equation}
	\label{equ13}
	\begin{split}
		Z = X + Y &= -r^n + \sum_{i=0}^{n-1} x_i r^i - r^n + \sum_{i=0}^{n-1} y_i r^i 
		= -r^{n+1} + \sum_{i=0}^{n} z_i r^i
	\end{split}
\end{equation}

To express the result in standard $(n+1)$-digit MRUD form:
\begin{equation}
	\label{equ14}
	\begin{split}
		Z &= -r^{n+1} + r^{n+1} - 2r^n + \sum_{i=0}^{n-1} (x_i + y_i) r^i \\
		&= -r^{n+1} + (r-2) r^n + \sum_{i=0}^{n-1} (x_i + y_i) r^i
	\end{split}
\end{equation}

Fig.~\ref{fig-add-sub}(a) illustrates the addition of two 2-digit radix-16 MRUD operands with digit set \([-16, 15]\).

\subsection{Subtraction Operation}\label{sec:sub_op}

The subtraction of two signed MRUD numbers is expressed as:
\begin{equation}
	\label{equ15}
	\begin{split}
		Z = X - Y &= -r^n + \sum_{i=0}^{n-1} x_i r^i + r^n - \sum_{i=0}^{n-1} y_i r^i\\
		&= -r^{n+1} + \sum_{i=0}^{n} z_i r^i
	\end{split}
\end{equation}

Using the digitwise complement $y_i = \beta - \overline{y_i}$:
\begin{equation}
	\label{equ16}
	\begin{split}
		Z &= -r^{n+1} + r^{n+1} - \sum_{i=0}^{n-1} y_i r^i + \sum_{i=0}^{n-1} x_i r^i 
		= -r^{n+1} + r^{n+1} - \sum_{i=0}^{n-1} (\beta - \overline{y_i}) r^i + \sum_{i=0}^{n-1} x_i r^i \\
		&= -r^{n+1} + r^{n+1} - \beta \sum_{i=0}^{n-1} r^i + \sum_{i=0}^{n-1} (x_i + \overline{y_i}) r^i \\
		&= -r^{n+1} + r^{n+1} - \beta \cdot \frac{r^n - 1}{r-1} + \sum_{i=0}^{n-1} (x_i + \overline{y_i}) r^i
	\end{split}
\end{equation}

For MRUD with $\beta = 2(r-1)+1$, this simplifies to:
\begin{equation}
	\label{equ17}
	Z = -r^{n+1} + (r-3) r^n + (r-2) \sum_{i=0}^{n-1} r^i + 3 + \sum_{i=0}^{n-1} (x_i + \overline{y_i}) r^i
\end{equation}

\begin{figure}[pos=htbp]
	\centering
	\scriptsize
	\begin{tabular}{ccc}
		\includegraphics[width=0.45\columnwidth, keepaspectratio=true]{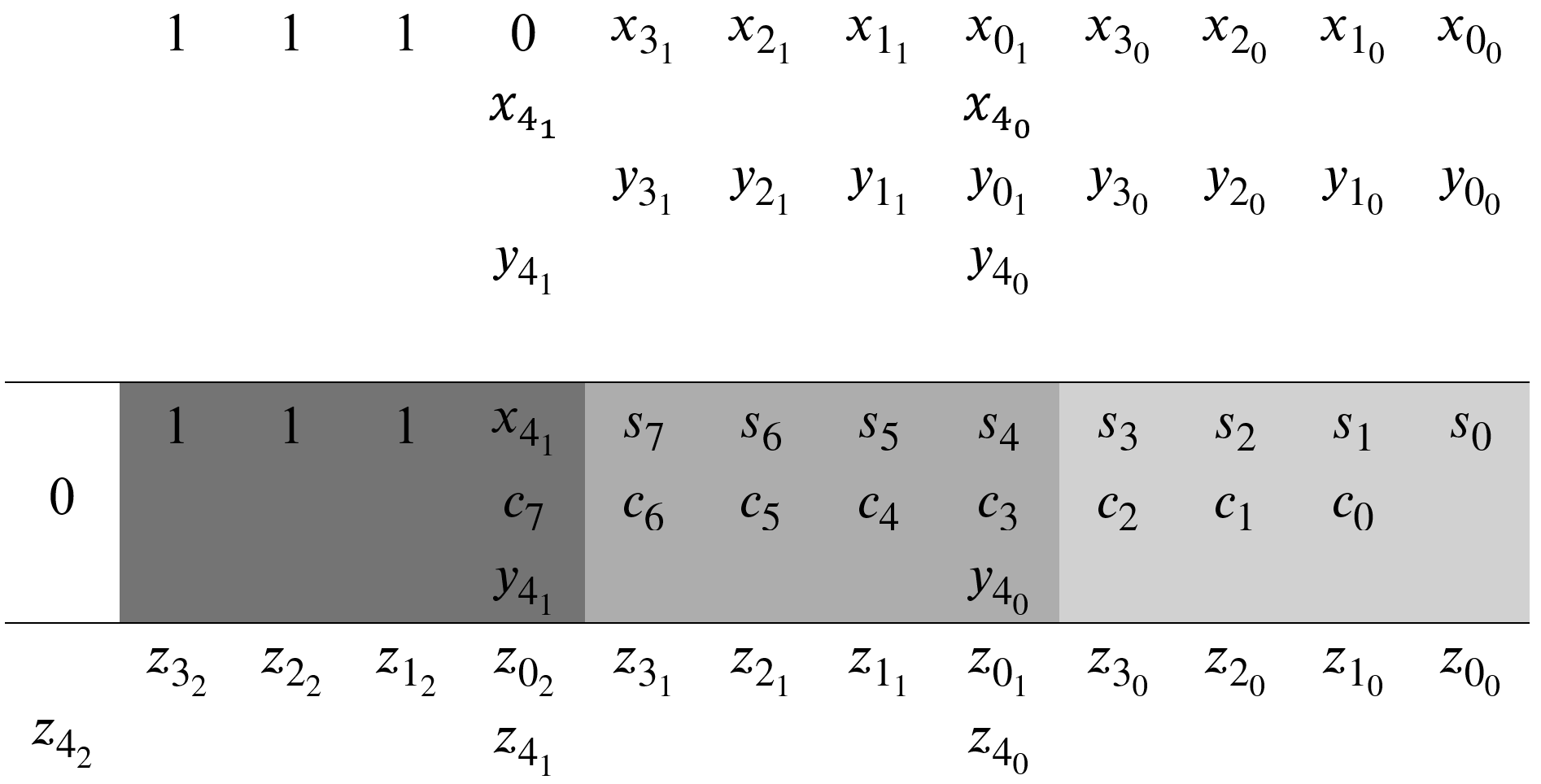} &
		\includegraphics[width=0.45\columnwidth, keepaspectratio=true]{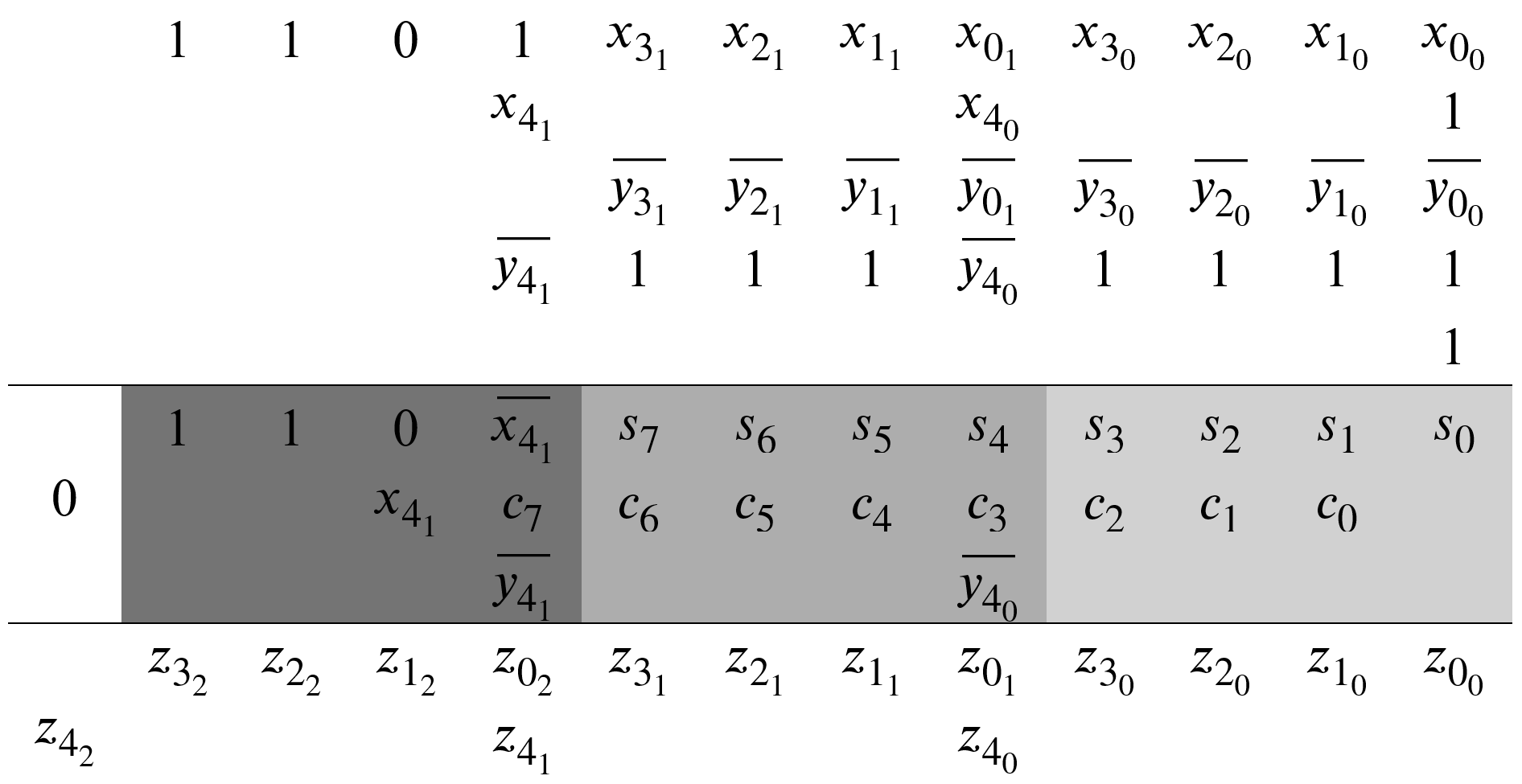} \\
		(a) Addition. & (b) Subtraction.
	\end{tabular}
	\caption{Radix-16 MRUD addition and subtraction operations.}
	\label{fig-add-sub}
\end{figure}

\begin{figure}[pos=htbp]
	\centering
	\scriptsize
		\includegraphics[width=0.5\columnwidth, keepaspectratio=true]{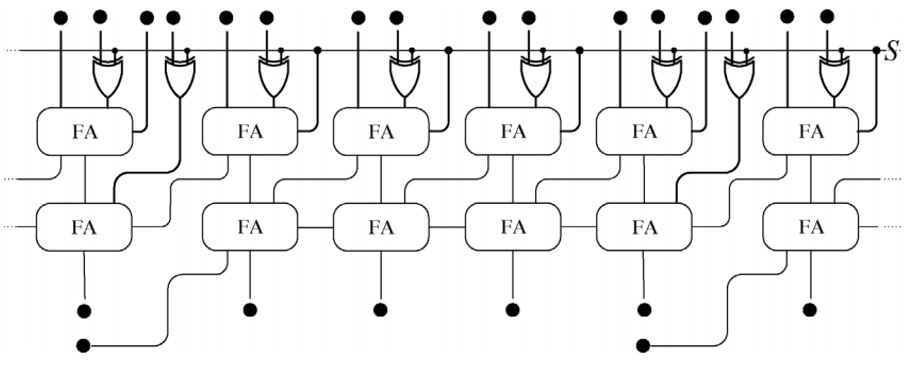} \\
	\caption{Single-digit slice of the unified adder/subtractor unit.}
	\label{fig-add-sub1}
\end{figure}

Fig.~\ref{fig-add-sub}(b) illustrates the subtraction of two 2-digit radix-16 MRUD operands with digit set \([-16, 15]\).
Based on Fig.~\ref{fig-add-sub}(a) and (b), a unified adder/subtractor can be derived, as shown in Fig.~\ref{fig-add-sub1}. This figure presents a single-digit slice of a redundant-digit radix-16 unified adder/subtractor unit, which uses a control signal $S$ to select addition ($S=0$) or subtraction ($S=1$).

\subsection{Overflow and Underflow Detection in Signed MRUD}

In non-redundant addition, overflow detection is straightforward: the carry-out from the MSD indicates overflow. However, in redundant addition, a non-zero carry digit does not necessarily indicate real overflow, since the combined value of the result digit and the carry may still fall within the valid range of the RDNS representation \citep{Jaberipur2017}.

As discussed in Section~\ref{sec:digit_extension}, a digit equal to $(r-1)$ can be eliminated from the most significant position. In the addition/subtraction operations shown in Fig.~\ref{fig-add-sub}(a) and (b), the third digit of the result ($Z_3$) can take the following values:
\begin{itemize}
	\item \textbf{Addition:} $(r-2)$, $(r-1)$, $r$, and $(r+1)$
	\item \textbf{Subtraction:} $(r-3)$, $(r-2)$, $(r-1)$, and $r$
\end{itemize}

When $Z_3 = (r-1)$, the MSD can be eliminated, resulting in an $n$-digit result. Otherwise, overflow/underflow occurs. However, if $Z_3$ can be transformed to $(r-1)$ through bit manipulation, the result can still be represented in $n$ digits. This condition is termed \textit{apparent overflow/underflow}. If such transformation is not possible, \textit{real overflow/underflow} occurs.

Table~\ref{tab:overflow_conditions} lists all possible values of the $(n+1)$-th digit ($Z_n$) along with the MSB of the $n$-th digit ($z_{4(n-1)}$). In cases where apparent overflow/underflow occurs, inverting the bit $z_{4(n-1)}$ yields the corrected value $z'_{4(n-1)}$, allowing $Z_n$ to be converted to $(r-1)$ and subsequently eliminated.

\begin{table*}[htbp]
	\centering
	\scriptsize
	\renewcommand{\arraystretch}{1.2}
	\caption{Possible overflow/underflow conditions for $n$-digit signed MRUD addition/subtraction with radix $r = 2^h$.}
	\label{tab:overflow_conditions}
	\begin{tabular}{ccccccccccccc}
		\toprule
		\multirow{2}{*}{\textit{Value of $Z_n$}} & \multicolumn{5}{c}{$Z_n (h=4)$} & \multirow{2}{*}{MSB of $Z_{n-1}$} & \multicolumn{2}{c}{Overflow in Addition} & \multicolumn{2}{c}{Underflow in Subtraction} & \multirow{2}{*}{Description} \\
		\cmidrule(lr){2-6} \cmidrule(lr){8-11}
		& $z_{4_n}$ & $z_{3_n}$ & $z_{2_n}$ & $z_{1_n}$ & $z_{0_n}$ & & Real & Apparent & Real & Apparent & \\
		\midrule
		\multirow{2}{*}{$(r-3)$}
		& 0 & 1 & 1 & 0 & 1 & 0 & X & X & X & X & Real Underflow \\
		& 0 & 1 & 1 & 0 & 1 & 1 & X & X & 1 & 0 & Real Underflow \\
		\midrule
		\multirow{2}{*}{$(r-2)$}
		& 0 & 1 & 1 & 1 & 0 & 0 & 1 & 0 & 1 & 0 & Real Overflow/Underflow \\
		& 0 & 1 & 1 & 1 & 0 & 1 & 0 & 1 & 0 & 1 & Apparent Overflow/Underflow \\
		\midrule
		\multirow{2}{*}{$(r-1)$}
		& 0 & 1 & 1 & 1 & 1 & 0 & 0 & 0 & 0 & 0 & No Overflow/Underflow \\
		& 0 & 1 & 1 & 1 & 1 & 1 & 0 & 0 & 0 & 0 & No Overflow/Underflow \\
		\midrule
		\multirow{2}{*}{$r$}
		& 1 & 0 & 0 & 0 & 0 & 0 & 0 & 1 & 0 & 1 & Apparent Overflow/Underflow \\
		& 1 & 0 & 0 & 0 & 0 & 1 & 1 & 0 & 1 & 0 & Real Overflow/Underflow \\
		\midrule
		\multirow{2}{*}{$(r+1)$}
		& 1 & 0 & 0 & 0 & 1 & 0 & 1 & 0 & X & X & Real Overflow \\
		& 1 & 0 & 0 & 0 & 1 & 1 & X & X & X & X & Real Overflow \\
		\bottomrule
	\end{tabular}
\end{table*}

From Table~\ref{tab:overflow_conditions}, the following logic expressions are derived for overflow/underflow detection:

\begin{equation}
	\label{equ18}
	\text{Real-Overflow/Underflow} = z_{1_n} \oplus (z_{0_n} \lor z_{h(n-1)})
\end{equation}
\begin{equation}
	\label{equ19}
	\text{Apparent-Overflow/Underflow} = \overline{z_{0_n}} \land (z_{1_n} \odot z_{h(n-1)})
\end{equation}
\begin{equation}
	\label{equ20}
	\begin{split}
		z'_{4(n-1)} &= \text{Apparent-Overflow/Underflow} \oplus z_{h(n-1)} \\
		&= [\overline{z_{0_n}} \land (z_{1_n} \odot z_{h(n-1)})] \oplus z_{h(n-1)}
	\end{split}
\end{equation}

\section{Multiplication in MRUD Number Systems}\label{sec:Multiplication}
MRSD-based digital multipliers suffer from high power consumption and area overhead. This section introduces multiplication operations for unsigned and signed MRUD number systems, where the all-positive digit representation reduces power consumption and area overhead.

\subsection{Digit-Wise Multiplication: MRSD vs. MRUD}
\begin{figure}[pos=htbp]
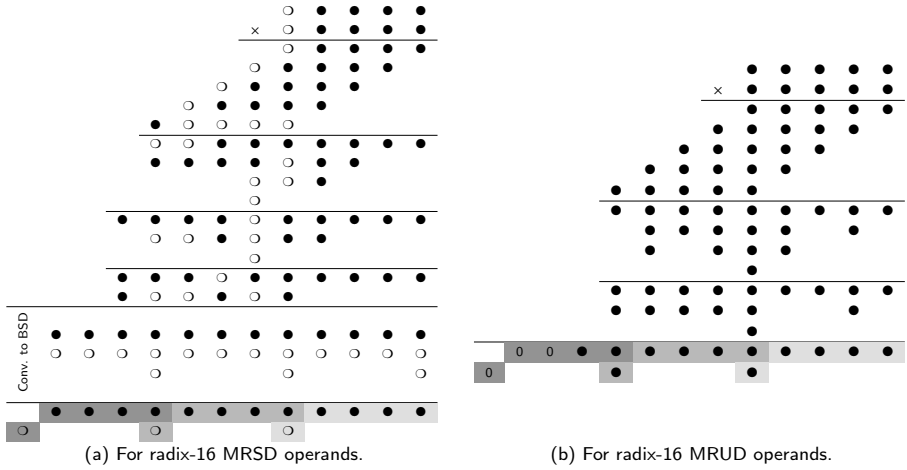

	\centering
	\scriptsize
	\begin{tabular}{cc}
		
		\resizebox{0.35\textwidth}{!}{%
			\begin{tabular}{ccccccccccccc}
				&&&&&&&& \ding{109} & \ding{108} & \ding{108} & \ding{108} & \ding{108}\\
				&&&&&&& \(\times\) & \ding{109} & \ding{108} & \ding{108} & \ding{108} & \ding{108}\\
				\cline{8-13}
				&&&&&&&& \ding{109} & \ding{108} & \ding{108} & \ding{108} & \ding{108}\\
				&&&&&&& \ding{109} & \ding{108} & \ding{108} & \ding{108} & \ding{108} &\\
				&&&&&& \ding{109} & \ding{108} & \ding{108} & \ding{108} & \ding{108} &&\\
				&&&&& \ding{109} & \ding{108} & \ding{108} & \ding{108} & \ding{108} &&&\\
				&&&& \ding{108} & \ding{109} & \ding{109} & \ding{109} & \ding{109} &&&&\\
				\cline{5-13}
				&&&& \ding{109} & \ding{109} & \ding{108} & \ding{108} & \ding{108} & \ding{108} & \ding{108} & \ding{108} & \ding{108}\\
				&&&& \ding{108} & \ding{108} & \ding{108} & \ding{108} & \ding{109} & \ding{108} & \ding{108} &&\\
				&&&&&&& \ding{109} & \ding{109} & \ding{108} &&&\\
				&&&&&&& \ding{109} &&&&&\\
				\cline{4-13}
				&&& \ding{108} & \ding{108} & \ding{108} & \ding{108} & \ding{109} & \ding{108} & \ding{108} & \ding{108} & \ding{108} & \ding{108} \\
				&&&& \ding{109} & \ding{109} & \ding{108} & \ding{109} & \ding{108} & \ding{108} &&&\\
				&&&&&&& \ding{109} &&&&&\\
				\cline{4-13}
				&&& \ding{108} & \ding{108} & \ding{108} & \ding{109} & \ding{108} & \ding{108} & \ding{108} & \ding{108} & \ding{108} & \ding{108} \\
				&&& \ding{108} & \ding{109} & \ding{109} & \ding{108} & \ding{109} & \ding{108} &&&&\\
				\hline
				\multirow{5}{*}{\rotatebox{90}{Conv. to BSD}}&&&&&&&&&&&&\\
				& \ding{108} & \ding{108} & \ding{108} & \ding{108} & \ding{108} & \ding{108} & \ding{108} & \ding{108} & \ding{108} & \ding{108} & \ding{108} & \ding{108} \\
				& \ding{109} & \ding{109} & \ding{109} & \ding{109} & \ding{109} & \ding{109} & \ding{109} & \ding{109} & \ding{109} & \ding{109} & \ding{109} & \ding{109} \\
				&&&& \ding{109} &&&& \ding{109} &&&& \ding{109} \\
				&&&&&&&&&&&&\\
				\hline
				& \cellcolor{gray!85}\ding{108} & \cellcolor{gray!85}\ding{108} & \cellcolor{gray!85}\ding{108} & \cellcolor{gray!85}\ding{108} & \cellcolor{gray!55}\ding{108} & \cellcolor{gray!55}\ding{108} & \cellcolor{gray!55}\ding{108} & \cellcolor{gray!55}\ding{108} & \cellcolor{gray!25}\ding{108} & \cellcolor{gray!25}\ding{108} & \cellcolor{gray!25}\ding{108} & \cellcolor{gray!25}\ding{108} \\
				\cellcolor{gray!85}\ding{109} &&&& \cellcolor{gray!55}\ding{109} &&&& \cellcolor{gray!25}\ding{109} &&&&\\
			\end{tabular}
		} &
		\resizebox{0.35\textwidth}{!}{%
			\begin{tabular}{ccccccccccccc}
				&&&&&&&& \ding{108} & \ding{108} & \ding{108} & \ding{108} & \ding{108} \\
				&&&&&&&\(\times\)& \ding{108} & \ding{108} & \ding{108} & \ding{108} & \ding{108} \\
				\cline{8-13}
				&&&&&&&& \ding{108} & \ding{108} & \ding{108} & \ding{108} & \ding{108} \\
				&&&&&&& \ding{108} & \ding{108} & \ding{108} & \ding{108} & \ding{108} &\\
				&&&&&& \ding{108} & \ding{108} & \ding{108} & \ding{108} & \ding{108} &&\\
				&&&&& \ding{108} & \ding{108} & \ding{108} & \ding{108} & \ding{108} &&&\\
				&&&& \ding{108} & \ding{108} & \ding{108} & \ding{108} & \ding{108} &&&&\\
				\cline{5-13}
				&&&& \ding{108} & \ding{108} & \ding{108} & \ding{108} & \ding{108} & \ding{108} & \ding{108} & \ding{108} & \ding{108} \\
				&&&&& \ding{108} & \ding{108} & \ding{108} & \ding{108} & \ding{108} && \ding{108} &\\
				&&&&& \ding{108} && \ding{108} & \ding{108} & \ding{108} &&&\\
				&&&&&&&& \ding{108} &&&&\\
				\cline{5-13}
				&&&& \ding{108} & \ding{108} & \ding{108} & \ding{108} & \ding{108} & \ding{108} & \ding{108} & \ding{108} & \ding{108} \\
				&&&& \ding{108} & \ding{108} & \ding{108} & \ding{108} & \ding{108} &&& \ding{108} &\\
				&&&&&&&& \ding{108} &&&&\\
				\hline
				& \cellcolor{gray!85}0 & \cellcolor{gray!85}0 & \cellcolor{gray!85}\ding{108} & \cellcolor{gray!85}\ding{108} & \cellcolor{gray!55}\ding{108} & \cellcolor{gray!55}\ding{108} & \cellcolor{gray!55}\ding{108} & \cellcolor{gray!55}\ding{108} & \cellcolor{gray!25}\ding{108} & \cellcolor{gray!25}\ding{108} & \cellcolor{gray!25}\ding{108} & \cellcolor{gray!25}\ding{108} \\
				\cellcolor{gray!85}0 &&&& \cellcolor{gray!55}\ding{108} &&&& \cellcolor{gray!25}\ding{108} &&&&\\
			\end{tabular}
		}
		\\
		
		(a) For radix-16 MRSD operands. & 		(b) For radix-16 MRUD operands.

	\end{tabular}
	
	\caption{Digit-wise multiplication of two RDNS operands in radix-16.}
	\label{fig_digit-wise-mul}
\end{figure}

In the multiplication of two \(n\)-digit MRUD numbers, the absence of negabits enables the efficient generation of partial products via digit-wise multiplication using binary sub-multipliers. Fig.~\ref{fig_digit-wise-mul} illustrates a comparative view of MRSD and MRUD sub-multipliers for radix-16 operands.

The MRSD sub-multiplier shown in Fig.~\ref{fig_digit-wise-mul}(a) employs a matrix of logic gates, including (\ding{108} \(\land\) \ding{108}), \(\overline{\text{\ding{109} \(\land\) \ding{109}}}\), and (\(\overline{\text{\ding{108}}}\) \(\lor\) \ding{109}), to produce the partial products. To ensure compatibility with the MRSD operand format, an additional transformation stage is required: the output is first converted into the BSD form, followed by conversion into MRSD representation.

In contrast, the MRUD sub-multiplier shown in Fig.~\ref{fig_digit-wise-mul}(b) generates partial products directly through a simple matrix of AND gates, since the representation uses only posibits. During the final accumulation stage, \(h\)-bit adders (with \(h=4\) for radix-16) are used to re-encode the sub-multiplier outputs into the input operand format.

These sub-multipliers operate fully in parallel. For radix-16 MRUD inputs, each sub-multiplier produces three MRUD partial product digits, which are then accumulated using carry-free addition---a key advantage of redundant number systems.

While MRSD inherently supports only signed multiplication, MRUD enables two distinct multiplier designs: one for unsigned operands and another for signed operands, as detailed in the following subsections.

\subsection{Signed MRUD Multiplication Algorithm}
Similar to the Baugh-Wooley multiplication algorithm \citep{baugh-wooley1973}, a parallel multiplication algorithm for signed MRUD number systems can be formulated based on the operands \(X\) and \(Y\) defined in \eqref{equ11} and \eqref{equ12}. Unlike conventional number systems, redundant multiplication may result in overflow. Therefore, the product is initially assumed to have \(2n+1\) digits. The product \(P = XY\) is expressed as:
\begin{equation}
	\label{equ21}
	\begin{split}
		P &= X \cdot Y 
		= \left( -r^n + \sum_{i=0}^{n-1} x_i r^i \right) \left( -r^n + \sum_{j=0}^{n-1} y_j r^j \right) \\
		&= -r^{2n+1} + \sum_{i=0}^{2n} p_i r^i
	\end{split}
\end{equation}

Let \(x_i = \beta - \overline{x_i}\) denote the digitwise complement of \(x_i\), and similarly for \(y_j\). Expanding \(P = XY\) using \eqref{equ11} and \eqref{equ12} yields:
\begin{equation}
	\label{equ22}
	\begin{split}
		P &= r^{2n} - r^n \sum_{j=0}^{n-1} y_j r^j - r^n \sum_{i=0}^{n-1} x_i r^i + \sum_{i=0}^{n-1}\sum_{j=0}^{n-1} x_i y_j r^{i+j} 
		= r^{2n} - r^n \sum_{j=0}^{n-1} (\beta - \overline{y_j}) r^j - r^n \sum_{i=0}^{n-1} (\beta - \overline{x_i}) r^i \\
		&\quad + \sum_{i=0}^{n-1}\sum_{j=0}^{n-1} x_i y_j r^{i+j}
	\end{split}
\end{equation}

After simplification, let \(C\) denote the correction digits:
\begin{equation}
	\label{equ23}
	C = r^{2n+1} + r^{2n} - 2\beta r^n \cdot \frac{r^n - 1}{r-1}
\end{equation}
Then, the product can be expressed as:
\begin{equation}
	\label{equ24}
	P = -r^{2n+1} + C - r^n \sum_{i=0}^{n-1} (\overline{x_i} + \overline{y_i}) r^i + \sum_{i=0}^{n-1}\sum_{j=0}^{n-1} x_i y_j r^{i+j}
\end{equation}

For the MRUD with \(\beta = 2(r-1) + 1\), the correction digits simplify to:
\begin{equation}
	\label{equ25}
	\begin{split}
		C &= r^{2n+1} + r^{2n} - 2(2r-1)r^n \cdot \frac{r^n - 1}{r-1} \\
		&= (r-4) r^{2n} + (r-3) \sum_{i=n+2}^{2n-1} r^i + (r-2) r^{n+1} + 2r^n
	\end{split}
\end{equation}

Fig.~\ref{fig:mult_matrix} illustrates the partial-product matrix for multiplying two \(n\)-digit signed MRUD operands with \(D=[0,\beta]\) and \(\beta = 2(r-1) + 1\). The elements of this matrix are generated by MRUD sub-multipliers (as shown in Fig.~\ref{fig_digit-wise-mul}(b) for radix-16), which produce three MRUD digits for radices greater than 2. The generated partial products, which are in MRUD format, are then accumulated using carry-free MRUD adders to produce the final product.

\begin{figure*}[pos=htbp]
	\centering
	\scriptsize
	\renewcommand{\arraystretch}{1.5}
	
	\begin{tabular}{ccccccccccccc}
		\hline
		Weights & $r^{2n}$ & $r^{2n-1}$ & $r^{2n-2}$ & $\cdot$ & $r^{n+1}$ & $r^n$ & $r^{n-1}$ & $r^{n-2}$ & $\cdots$ & $r^2$ & $r^1$ & $r^0$ \\
		\hline
		& & & & & & & $X_{n-1}$ & $X_{n-2}$ & $\cdots$ & $X_2$ & $X_1$ & $X_0$ \\
		& & & & & & $\times$ & $Y_{n-1}$ & $Y_{n-2}$ & $\cdots$ & $Y_2$ & $Y_1$ & $Y_0$ \\
		\cline{8-13}
		& & & & & & $\overline{Y_0}$ & $X_{n-1}Y_0$ & $X_{n-2}Y_0$ & $\cdots$ & $X_2Y_0$ & $X_1Y_0$ & $X_0Y_0$ \\
		& & & & & $\overline{Y_1}$ & $X_{n-1}Y_1$ & $X_{n-2}Y_1$ & $\cdots$ & $X_2Y_1$ & $X_1Y_1$ & $X_0Y_1$ & \\
		& & & & $\overline{Y_2}$ & $X_{n-1}Y_2$ & $X_{n-2}Y_2$ & $\cdots$ & $X_2Y_2$ & $X_1Y_2$ & $X_0Y_2$ & & \\
		& & & $\iddots$ & $\iddots$ & $\iddots$ & $\cdots$ & $\iddots$ & $\iddots$ & $\iddots$ & & & \\
		& & $\overline{Y_{n-2}}$ & $X_{n-1}Y_{n-2}$ & $X_{n-2}Y_{n-2}$ & $\cdots$ & $X_2Y_{n-2}$ & $X_1Y_{n-2}$ & $X_0Y_{n-2}$ & & & & \\
		& $\overline{Y_{n-1}}$ & $X_{n-1}Y_{n-1}$ & $X_{n-2}Y_{n-1}$ & $\cdots$ & $X_2Y_{n-1}$ & $X_1Y_{n-1}$ & $X_0Y_{n-1}$ & & & & & \\
		& $\overline{X_{n-1}}$ & $\overline{X_{n-2}}$ & $\cdots$ & $\overline{X_2}$ & $\overline{X_1}$ & $\overline{X_0}$ & & & & & & \\
		& $(r-4)$ & $(r-3)$ & $\cdots$ & $(r-3)$ & $(r-2)$ & $2$ & \multicolumn{2}{c}{\textcolor{red}{\textbf{$\leftarrow$ Correction Digits}}} & & & & \\
		\hline
		& $\vdots$ & & & & & $\vdots$ & & & & & & $\vdots$ \\
		\hline
		& $P_{2n}$ & $P_{2n-1}$ & $P_{2n-2}$ & $\cdots$ & $P_{n+1}$ & $P_{n}$ & $P_{n-1}$ & $P_{n-2}$ & $\cdots$ & $P_2$ & $P_1$ & $P_0$ \\
	\end{tabular}
	\caption{Partial-product array for signed $n$-digit MRUD multiplication with $D=[0,\beta]$ and $\beta=2(r-1)+1$.}
	\label{fig:mult_matrix}
\end{figure*}

As observed in Fig.~\ref{fig:mult_matrix}, the final product has \(2n+1\) digits. Using the overflow/underflow detection scheme presented in \eqref{equ18}, \eqref{equ19}, and \eqref{equ20} (Section~\ref{sec:Unified}), real and apparent overflow conditions can be identified. In the case of apparent overflow, the last bit of the \(n\)-th digit of the final product can be corrected, allowing the final product to be represented in \(n\) digits. Since both the input operands and the final product are in redundant MRUD format, this multiplier is referred to as the \textit{RRR-MRUD} (Redundant inputs, Redundant output) multiplier. In RRR-MRUD multiplication, the correction digits occupy \(n+1\) positions.

\begin{figure}[pos=htbp]
	\centering
	\scriptsize
	\includegraphics[width=0.65\columnwidth, keepaspectratio=true]{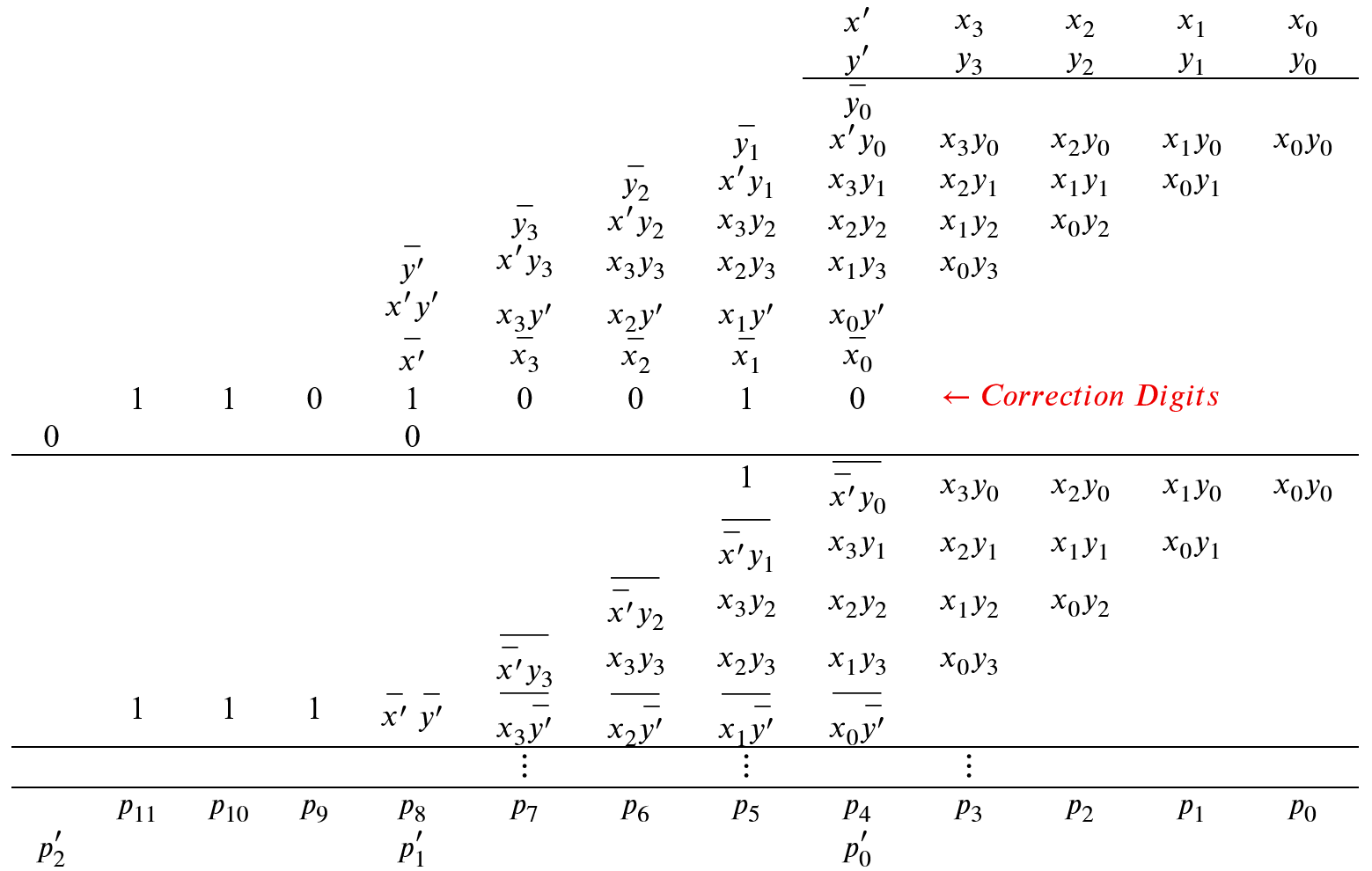}
	\caption{Partial product generation and correction terms for $1\times1$ RRR-MRUD.}
	\label{fig:ppg_1x1_rrr}
\end{figure}

\begin{figure}[pos=htbp]
	\centering
	\scriptsize
	\includegraphics[width=0.9\columnwidth, keepaspectratio=true]{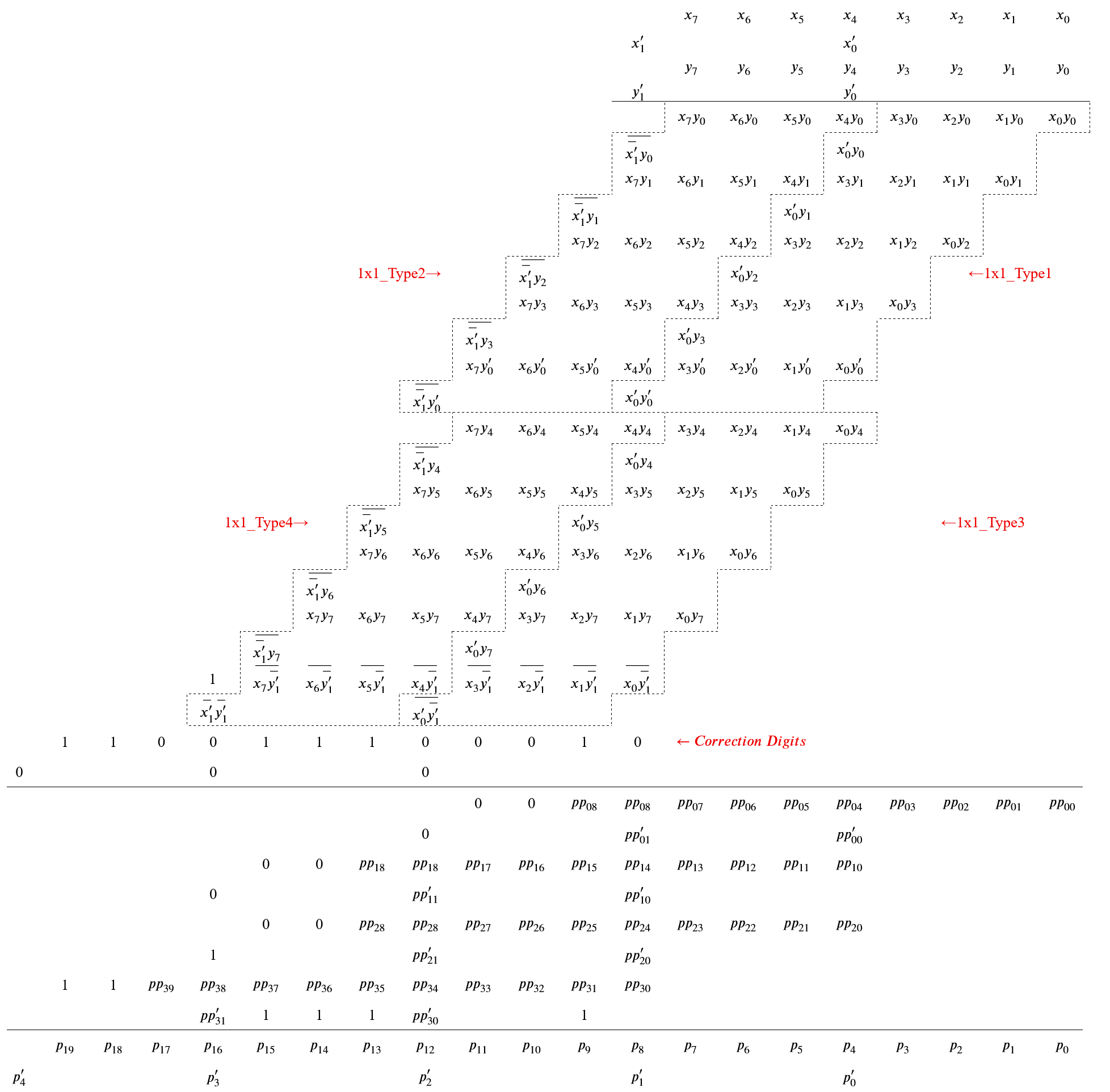}
	\caption{Partial product generation and correction terms for $2\times2$ RRR-MRUD.}
	\label{fig:ppg_2x2_rrr}
\end{figure}

Fig.~\ref{fig:ppg_1x1_rrr} shows a $1\times1$ RRR multiplier, where two 1-digit MRUD operands generate a 3-digit product (\(2n+1 = 3\)). The partial product matrix and its reduction stages are illustrated. Fig.~\ref{fig:ppg_2x2_rrr} extends this to $2\times2$ RRR multiplication, where four $1\times1$ sub-multipliers generate partial products, which are then combined using MRUD adders.

Alternatively, when the inputs are converted from binary to MRUD and then multiplied using the MRUD multiplier, the output is an \(n\)-digit redundant product in MRUD format. This configuration is referred to as the \textit{BBR-MRUD} (Binary inputs, Redundant output) multiplier.

\begin{figure}[pos=htbp]
	\centering
	\scriptsize
	\begin{tabular}{c}
		\includegraphics[width=0.5\columnwidth, keepaspectratio=true]{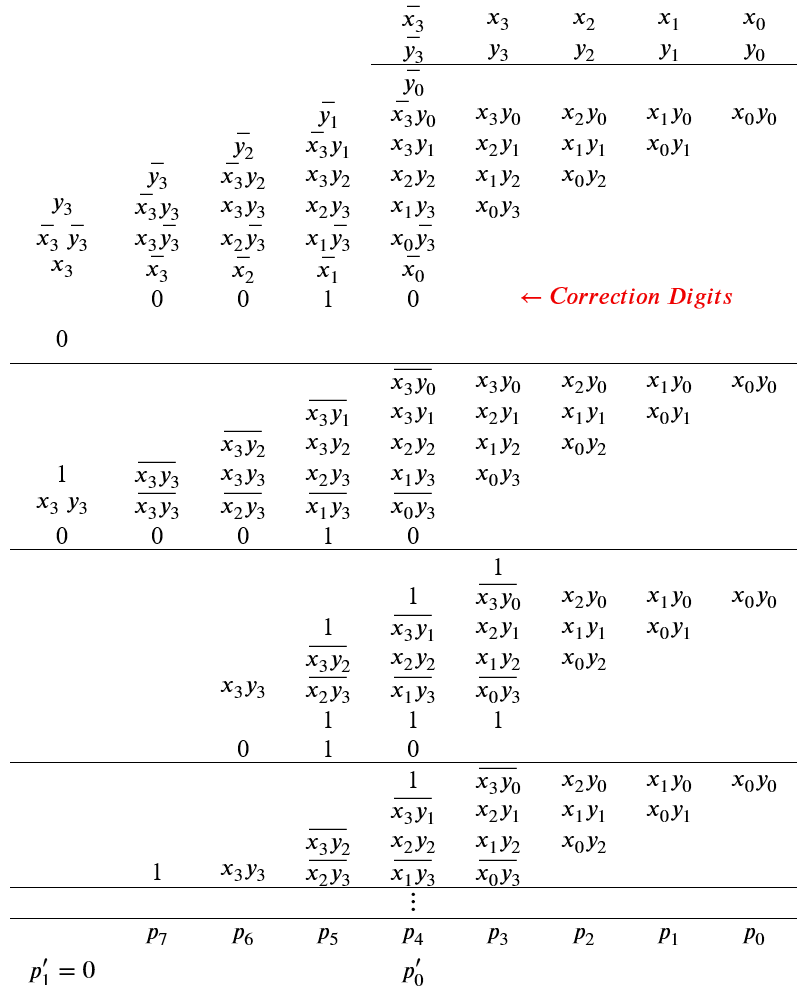}\\
		(a) For $1\times1$ BBR-MRUD operands.\\
		\includegraphics[width=0.9\columnwidth, keepaspectratio=true]{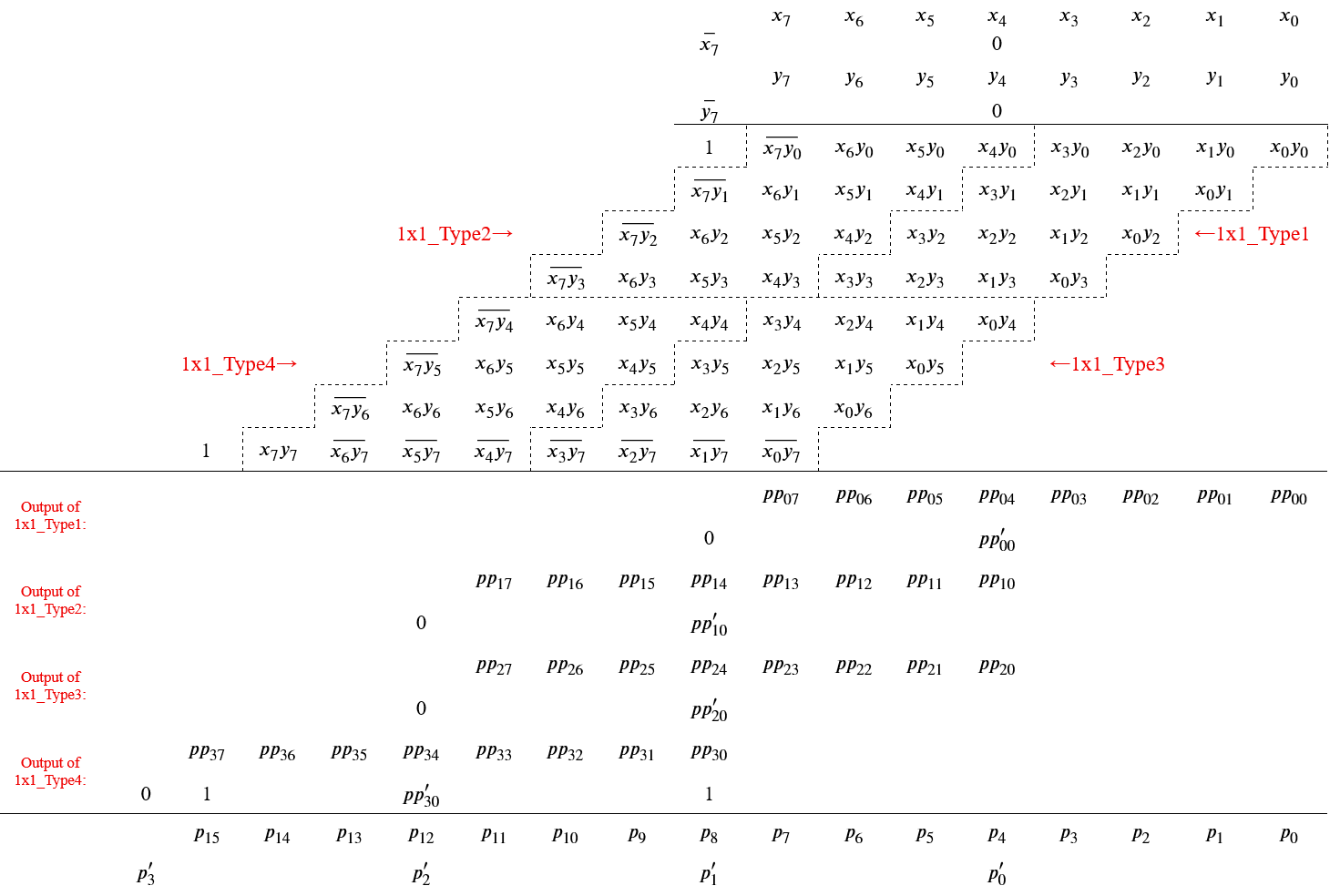} \\
		(b) For $2\times2$ BBR-MRUD operands.
	\end{tabular}
	\caption{Partial product generation and correction terms for BBR multipliers.}
	\label{fig:ppg_bbr}
\end{figure}

Fig.~\ref{fig:ppg_bbr}(a) shows the simplified partial-product matrix for a $1\times1$-digit BBR-MRUD multiplier. The figure shows the logical simplification stages leading to the final reduced partial product matrix, which resembles the well-known Baugh-Wooley algorithm. In BBR-MRUD multiplication, the correction digits occupy \(n\) positions. Fig.~\ref{fig:ppg_bbr}(b) extends this approach to $2\times2$ multiplication, where each operand is converted to two 2-digit MRUD numbers. This multiplier consists of four $1\times1$ sub-multipliers, each generating a partial product. The partial products are then combined using MRUD adders to produce a 4-digit redundant output.

Both RRR and BBR multipliers can be scaled hierarchically to support larger operand sizes. In both configurations, larger multipliers are constructed from smaller sub-multipliers: $2\times2$ multipliers use four $1\times1$ sub-multipliers, $4\times4$ multipliers use four $2\times2$ sub-multipliers, and $8\times8$ multipliers use four $4\times4$ sub-multipliers. The RRR configuration produces a \(2n+1\)-digit output, while the BBR configuration produces a \(2n\)-digit output.

\subsection{Unsigned MRUD Multiplication}

In unsigned MRUD multiplication, all digits are positive and there is no sign bit to handle. Consequently, the partial product generation and reduction stages are significantly simpler than in the signed case. The product of two \(n\)-digit unsigned MRUD operands is given by:
\begin{equation}
	\label{equ26}
	P = \sum_{i=0}^{n-1}\sum_{j=0}^{n-1} x_i y_j r^{i+j}, \quad 0 \leq x_i, y_j \leq \beta
\end{equation}

Unlike signed multiplication, unsigned MRUD multiplication does not require bitwise complementation and correction digits. This results in a more compact partial product matrix and lower hardware overhead.

Similar to signed multiplication, in unsigned MRUD multiplication, when redundant inputs are applied to the multiplier and a redundant output is produced (RRR-MRUD), overflow may occur. However, when the inputs are converted from binary to redundant and then applied to the multiplier (BBR-MRUD), the produced output is in redundant format with \(2n\) digits and no overflow occurs.

\begin{figure}[pos=htbp]
	\centering
	\scriptsize
	\begin{tabular}{cc}
		\includegraphics[width=0.55\columnwidth, keepaspectratio=true]{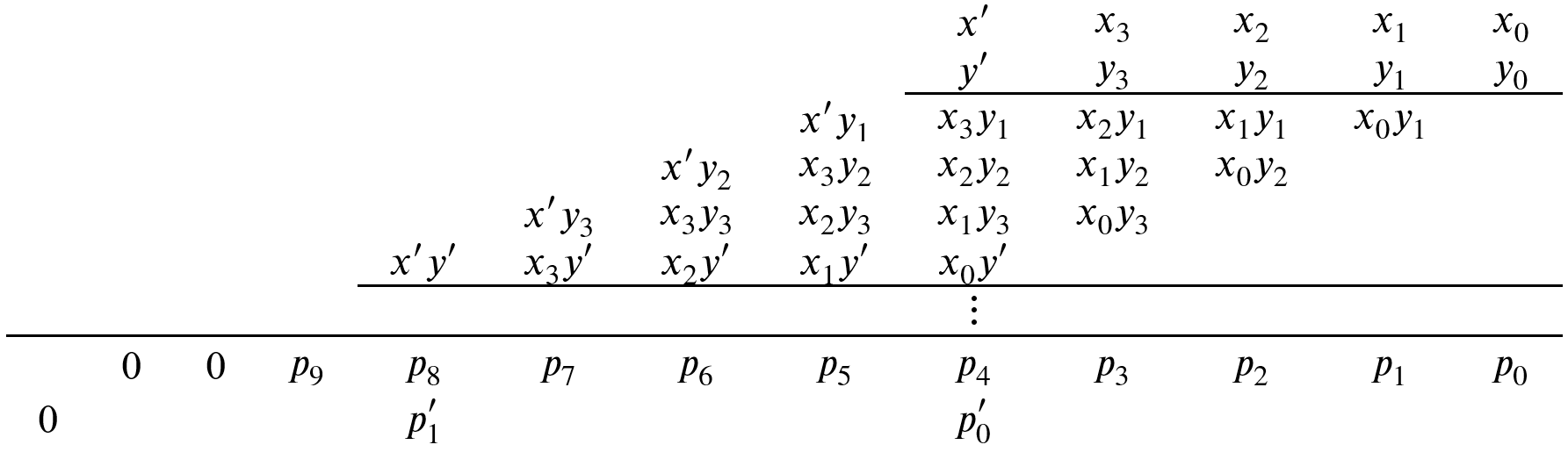}& 	\includegraphics[width=0.4\columnwidth, keepaspectratio=true]{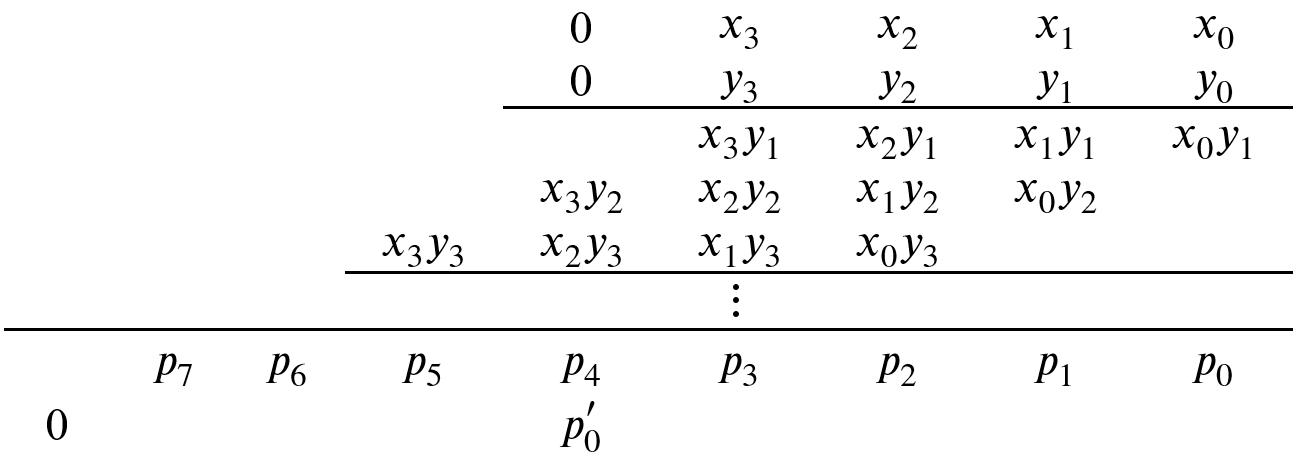}\\
		(a) $1\times1$ RRR unsigned multiplication.& (b) $1\times1$ BBR unsigned multiplication.\\	
	\end{tabular}
	\caption{Unsigned MRUD multiplication: (a) $1\times1$ RRR, (b) $1\times1$ BBR.}
	\label{fig:unsigned_mul}
\end{figure}

Fig.~\ref{fig:unsigned_mul}(a) illustrates a $1\times1$ RRR unsigned multiplication, where two 1-digit unsigned MRUD operands produce a \(2n+1 = 3\)-digit redundant product. As shown in the figure, no correction terms or inversions are needed. Fig.~\ref{fig:unsigned_mul}(b) shows a $1\times1$ BBR unsigned multiplication, where two 1-digit unsigned MRUD operands produce a \(2n = 2\)-digit redundant product.

The hierarchical scaling approach described for signed multipliers also applies to unsigned multipliers, enabling larger implementations using the same recursive structure.

\section{Experimental Results}\label{sec:Experimental_Results}

This section presents the synthesis and post-synthesis simulation results for the proposed MRUD-based multipliers. All designs were synthesized using Synopsys Design Compiler 2018 with the FreePDK 45nm library. Post-synthesis simulations were performed using ModelSim SE 2020.4 to generate SAIF activity files, which were then back-annotated to Design Compiler for accurate power estimation.

To determine the critical path delay of each design, we applied an iterative timing closure approach. Starting with a loose constraint, we gradually tightened the maximum delay until the design no longer met timing. The reported delay corresponds to the tightest achievable constraint with non-negative slack.

\subsection{RRR Multipliers: MRSD vs. MRUD}

Table~\ref{tab:rrr_results} compares the performance of signed MRSD, signed MRUD, and unsigned MRUD multipliers in RRR mode for operand sizes ranging from $1\times1$ to $8\times8$ digits in radix-16. The percentage improvements (or degradations) are shown relative to signed MRSD.

\begin{table}[htbp]
	\centering
	\tiny
	\setlength{\tabcolsep}{3pt}
	\caption{Performance comparison of RRR multipliers (MRSD \citep{amanollahi-AMR2022} vs. MRUD).}
	\label{tab:rrr_results}
	\begin{tabular}{cccccccc}
		\toprule
		\multirow{2}{*}{Size(digits)} & \multirow{2}{*}{Type} & \multirow{2}{*}{Arch.} & Delay & Freq & Area & Power & Energy \\
		& & & (ns) & (MHz) & (µm²) & (µW) & (fJ) \\
		\midrule
		\multirow{3}{*}{$1\times1$} & Signed & MRSD & 0.63 & 1587 & 137.79 & 47.64 & 30.01 \\
		& Signed & MRUD & 0.52($\downarrow$17\%) & 1923 & 138.59($\uparrow$1\%) & 54.58($\uparrow$15\%) & 28.38($\downarrow$5\%) \\
		& Unsigned & MRUD & 0.51($\downarrow$19\%) & 1961 & 115.44($\downarrow$16\%) & 19.71($\downarrow$59\%) & 10.05($\downarrow$67\%) \\
		\midrule
		\multirow{3}{*}{$2\times2$} & Signed & MRSD & 1.33 & 752 & 760.23 & 640.48 & 851.84 \\
		& Signed & MRUD & 0.93($\downarrow$30\%) & 1075 & 671.38($\downarrow$12\%) & 320.76($\downarrow$50\%) & 298.31($\downarrow$65\%) \\
		& Unsigned & MRUD & 0.96($\downarrow$28\%) & 1042 & 630.42($\downarrow$17\%) & 172.56($\downarrow$73\%) & 165.66($\downarrow$81\%) \\
		\midrule
		\multirow{3}{*}{$4\times4$} & Signed & MRSD & 1.74 & 575 & 3603.77 & 4380.70 & 7622.42 \\
		& Signed & MRUD & 1.41($\downarrow$19\%) & 709 & 2757.36($\downarrow$23\%) & 1607.90($\downarrow$63\%) & 2267.14($\downarrow$70\%) \\
		& Unsigned & MRUD & 1.38($\downarrow$21\%) & 725 & 2699.37($\downarrow$25\%) & 889.46($\downarrow$80\%) & 1227.45($\downarrow$84\%) \\
		\midrule
		\multirow{3}{*}{$8\times8$} & Signed & MRSD & 2.33 & 429 & 14621.22 & 31459.00 & 73299.47 \\
		& Signed & MRUD & 1.95($\downarrow$16\%) & 513 & 11308.99($\downarrow$23\%) & 6246.40($\downarrow$80\%) & 12180.48($\downarrow$83\%) \\
		& Unsigned & MRUD & 1.99($\downarrow$15\%) & 503 & 10800.13($\downarrow$26\%) & 3985.20($\downarrow$87\%) & 7930.55($\downarrow$89\%) \\
		\bottomrule
	\end{tabular}
\end{table}

As shown in Table~\ref{tab:rrr_results}, MRUD-based multipliers consistently outperform their MRSD counterparts across most metrics. The signed MRUD achieves 16--30\% lower delay and 12--23\% area reduction compared to MRSD, with power savings of 50--80\%. These improvements are primarily attributed to the simplified partial product generation stage in MRUD. Since MRUD uses only positive bits, complex mixed-polarity logic gates and the intermediate BSD conversion stage are eliminated. The unsigned MRUD offers further improvements, achieving up to 28\% delay reduction, 26\% area reduction, and 87\% power reduction compared to signed MRSD, as the complete removal of sign-related logic simplifies the overall architecture. For the $1\times1$ case, the signed MRUD shows a marginal area increase (1\%) and power increase (15\%), suggesting that the overhead of sign inversion logic becomes more significant at smaller sizes. The energy efficiency (PDP) of unsigned MRUD is particularly impressive, with 67--89\% reduction across all sizes, confirming that the structural simplifications in MRUD translate into tangible performance gains.

\subsection{BBR/BBB Multipliers: MRUD vs. Conventional}

Table~\ref{tab:signed_unsigned_combined} compares the performance of signed and unsigned MRUD multipliers (BBR and BBB modes) with conventional multiplier architectures, including DesignWare, Modified Booth, Dadda, and Wallace.

\begin{table*}[!htbp]
	\centering
	\tiny
	\setlength{\tabcolsep}{3pt}
	\caption{Comparison of signed and unsigned multipliers (BBR/BBB MRUD vs. conventional).}
	\label{tab:signed_unsigned_combined}
	\begin{tabular}{clrrrrrrrrrr}
		\toprule
		\multirow{2}{*}{Size} & \multirow{2}{*}{Architecture} & \multicolumn{5}{c}{\textbf{Signed}} & \multicolumn{5}{c}{\textbf{Unsigned}} \\
		\cmidrule(lr){3-7} \cmidrule(lr){8-12}
		& & \textbf{Delay (ns)} & \textbf{Freq (MHz)} & \textbf{Area (µm²)} & \textbf{Power (µW)} & \textbf{Energy (fJ)} & \textbf{Delay (ns)} & \textbf{Freq (MHz)} & \textbf{Area (µm²)} & \textbf{Power (µW)} & \textbf{Energy (fJ)} \\
		\midrule
		\multirow{6}{*}{$4\times4$} & DesignWare & 0.48 & 2083 & 72.62 & 109.76 & 52.68 & 0.42 & 2381 & 79.00 & 107.97 & 45.35 \\
		& Modified Booth & 0.54 & 1852 & 81.93 & 167.84 & 90.63 & --- & --- & --- & --- & --- \\
		& Dadda & 0.44 & 2273 & 82.19 & 155.59 & 68.46 & 0.47 & 2128 & 76.34 & 119.31 & 56.08 \\
		& Wallace & 0.44 & 2273 & 72.09 & 120.26 & 52.91 & 0.47 & 2128 & 79.27 & 123.09 & 57.85 \\
		& BBB-MRUD & 0.45 & 2222 & 67.03 & 123.11 & 55.40 & 0.46 & 2174 & \textbf{67.56} & \textbf{95.85} & 44.09 \\
		& BBR-MRUD & \textbf{0.33} & \textbf{3030} & \textbf{64.11} & \textbf{104.31} & \textbf{34.42} & \textbf{0.31} & \textbf{3226} & 71.29 & 115.05 & \textbf{35.66} \\
		\midrule
		\multirow{6}{*}{$8\times8$} & DesignWare & 0.90 & 1111 & \textbf{327.98} & 609.37 & 548.43 & 0.87 & 1149 & 392.62 & 590.54 & 513.77 \\
		& Modified Booth & 0.92 & 1087 & 343.41 & 684.47 & 629.71 & --- & --- & --- & --- & --- \\
		& Dadda & 0.89 & 1124 & 406.45 & 616.32 & 548.52 & 0.91 & 1099 & \textbf{355.11} & 515.92 & 469.49 \\
		& Wallace & 0.87 & 1149 & 415.23 & 606.97 & 528.06 & 0.86 & 1163 & 387.30 & 582.42 & 500.88 \\
		& BBB-MRUD & 0.93 & 1075 & 408.84 & 635.20 & 590.74 & 0.99 & 1010 & 406.18 & 601.01 & 595.00 \\
		& BBR-MRUD & \textbf{0.74} & \textbf{1351} & 348.46 & \textbf{580.94} & \textbf{429.90} & \textbf{0.74} & \textbf{1351} & 357.50 & \textbf{508.78} & \textbf{376.49} \\
		\midrule
		\multirow{6}{*}{$16\times16$} & DesignWare & 1.56 & 641 & 1427.36 & 1751.90 & 2732.96 & 1.57 & 637 & 1601.05 & 2032.20 & 3190.55 \\
		& Modified Booth & 1.58 & 633 & \textbf{1378.15} & 1876.60 & 2965.03 & --- & --- & --- & --- & --- \\
		& Dadda & 1.55 & 645 & 1659.57 & \textbf{1647.90} & 2554.25 & 1.67 & 599 & 1646.81 & \textbf{1570.70} & 2623.07 \\
		& Wallace & 1.68 & 595 & 1678.46 & 1779.20 & 2989.06 & 1.65 & 606 & 1725.81 & 1765.10 & 2912.42 \\
		& BBB-MRUD & 1.52 & 658 & 1743.36 & 2036.10 & 3094.87 & 1.49 & 671 & 1798.16 & 2030.70 & 3025.74 \\
		& BBR-MRUD & \textbf{1.15} & \textbf{870} & 1684.31 & 1988.80 & \textbf{2287.12} & \textbf{1.18} & \textbf{847} & \textbf{1568.07} & 1706.90 & \textbf{2014.14} \\
		\midrule
		\multirow{6}{*}{$32\times32$} & DesignWare & 4.42 & 226 & \textbf{4113.42} & \textbf{5797.80} & 25626.28 & 4.10 & 244 & \textbf{4456.83} & 5027.50 & 20612.75 \\
		& Modified Booth & 2.97 & 337 & 5050.54 & 7181.50 & 21329.06 & --- & --- & --- & --- & --- \\
		& Dadda & 2.87 & 348 & 6808.80 & 5931.20 & 17022.54 & 2.98 & 336 & 6568.07 & \textbf{4630.50} & 13798.89 \\
		& Wallace & 3.24 & 309 & 6448.37 & 6089.80 & 19730.95 & 3.25 & 308 & 6372.30 & 4904.40 & 15939.30 \\
		& BBB-MRUD & 2.56 & 391 & 7338.41 & 9294.00 & 23792.64 & 2.60 & 385 & 7233.87 & 6937.00 & 18036.20 \\
		& BBR-MRUD & \textbf{1.64} & \textbf{610} & 7039.42 & 8236.90 & \textbf{13508.52} & \textbf{1.64} & \textbf{610} & 6988.35 & 8375.70 & \textbf{13736.15} \\
		\bottomrule
	\end{tabular}
\end{table*}

As shown in Table~\ref{tab:signed_unsigned_combined}, BBR-MRUD achieves the lowest delay among all architectures for both signed and unsigned modes across all sizes. The main reason for this improvement is that BBR-MRUD maintains the result in redundant MRUD format, eliminating the carry-propagation overhead of the final binary conversion. In contrast, BBB-MRUD includes an additional reverse conversion stage, which makes its delay comparable to conventional multipliers. This confirms that the primary benefit of MRUD in multiplication is realized when the redundant output is directly used in subsequent redundant operations, avoiding the costly conversion to binary.

The unsigned variants consistently outperform their signed counterparts due to the elimination of sign-related operations, further validating the benefits of unsigned MRUD representations.

\begin{figure*}[pos=htbp]
	\centering
	\scriptsize
	\includegraphics[width=\columnwidth, keepaspectratio=true]{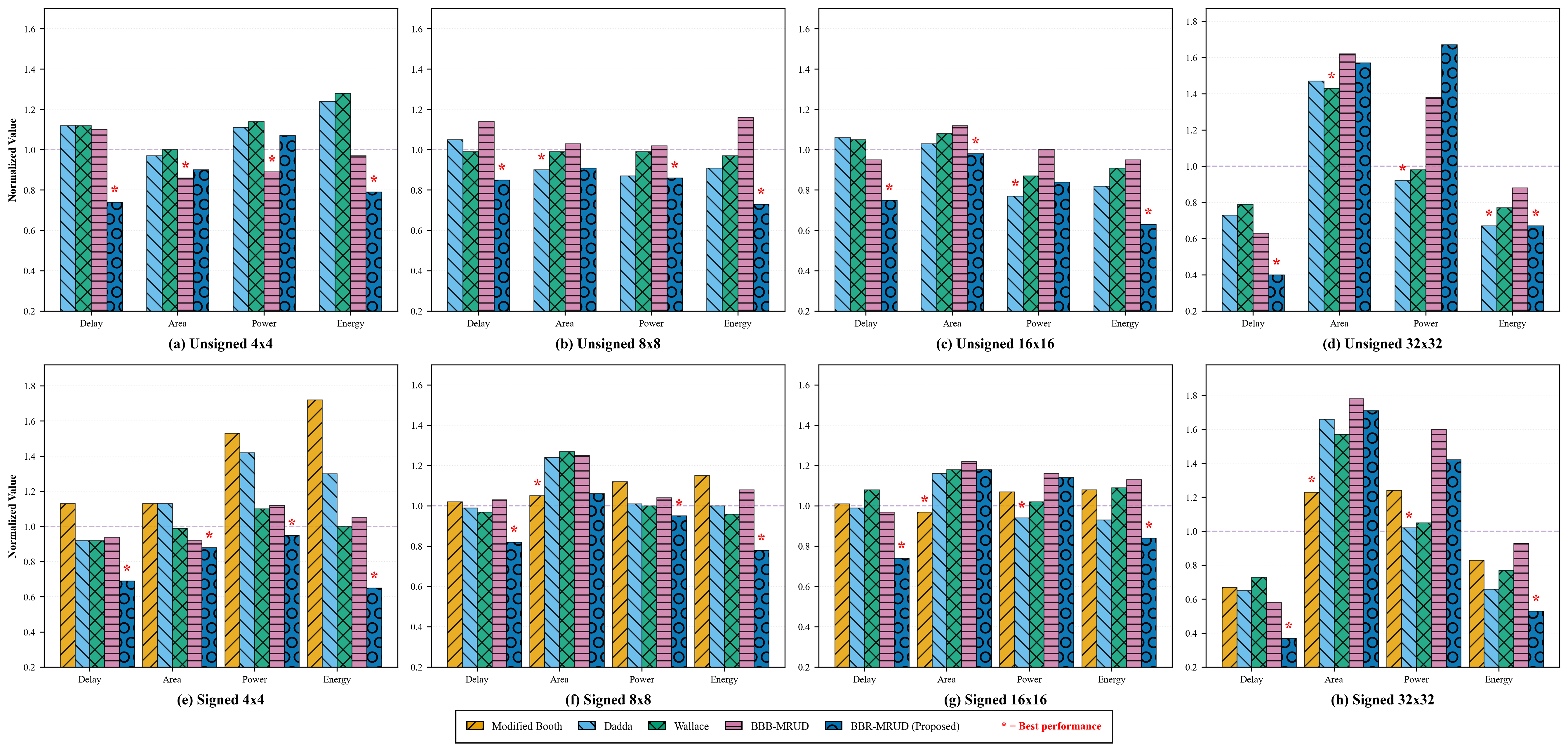}
	\caption{Normalized performance metrics for signed and unsigned multipliers across different sizes. All metrics are normalized to DesignWare.}
	\label{fig:normalized_metrics}
\end{figure*}

To facilitate a more intuitive comparison, Fig.~\ref{fig:normalized_metrics} presents the performance metrics normalized to the DesignWare implementation for each size. Fig.~\ref{fig:normalized_metrics}(a)--(d) show the normalized delay, area, power, and energy for unsigned multipliers across $4\times4$, $8\times8$, $16\times16$, and $32\times32$ sizes, respectively. Fig.~\ref{fig:normalized_metrics}(e)--(h) present the same metrics for signed multipliers. Lower values indicate better performance, and a value of 1.0 corresponds to the DesignWare baseline.

As observed in Fig.~\ref{fig:normalized_metrics}, BBR-MRUD consistently achieves the lowest normalized delay and energy across all sizes and both signed and unsigned modes. The unsigned variants show even greater improvements, with BBR-MRUD achieving up to 26\% lower delay and 21\% lower energy than DesignWare for $32\times32$. These visual results confirm the numerical trends presented in Table~\ref{tab:signed_unsigned_combined}, reinforcing the effectiveness of the MRUD representation for high-performance multiplier design.

\subsection{MAC Unit Evaluation}

To evaluate the effectiveness of the proposed MRUD-based MAC unit, four representative MAC architectures---DesignWare, Modified Booth, Dadda, and Wallace---were synthesized using Synopsys Design Compiler 2018 with the FreePDK 45nm library. The minimum achievable clock period was extracted from post-synthesis timing reports to determine the maximum operating frequency.

Post-synthesis simulations were performed in ModelSim SE 2020.4 using the generated gate-level netlist, the standard-cell Verilog library, and the back-annotated Standard Delay Format (SDF) file. All architectures were evaluated using the same set of 100 random input vectors to ensure a consistent and fair comparison. The simulation environment enforced realistic timing constraints, including setup and hold times derived from the synthesis reports, and measured the total execution time for all 100 operations. Throughput was then calculated as the number of operations per second based on this measured time. Power was estimated by generating SAIF activity files from the simulations and back-annotating them to Design Compiler.

As discussed in Section~\ref{sec:Multiplication}, the BBR mode maintains results in the redundant domain. In MAC units, this advantage is amplified because reverse conversion is performed only once at the end of the accumulation loop, eliminating carry propagation during intermediate additions.

Table~\ref{tab:mac_results} summarizes the synthesis and simulation results. The proposed MRUD-based MAC consistently achieves the shortest delay and highest throughput across all operand widths. For 32-bit operands, the MRUD MAC achieves a delay of 2.14 ns, representing a 40\% improvement over DesignWare (3.56 ns) and over 59\% improvement over Booth, Dadda, and Wallace. This translates to a maximum frequency of 461 MHz and a throughput of 152.6 MOPS, a 64\% improvement over DesignWare and approximately 2.5$\times$ higher than Wallace. The bandwidth similarly improves from 2.98 Gbps to 4.88 Gbps.

\begin{table*}[!t]
	\centering
	\caption{Synthesis and simulation results for sequential MAC architectures.}
	\label{tab:mac_results}
	\renewcommand{\arraystretch}{1.1}
	\tiny
	\setlength{\tabcolsep}{3pt}
	\begin{tabular}{clcccccccc}
		\toprule
		\multirow{2}{*}{Size(bits)} &
		\multirow{2}{*}{Architecture} &
		\multicolumn{4}{c}{\textbf{Synthesis Results}} &
		\multicolumn{4}{c}{\textbf{Simulation Results}} \\
		\cmidrule(lr){3-6} \cmidrule(lr){7-10}
		&
		&
		Delay
		(ns)
		&
		Area
		($\mu m^2$)
		&
		Power
		($\mu W$)
		&
		Energy
		(fJ)
		&
		Frequency
		(MHz)
		&
		Throughput
		(MOPS)
		&
		Time/Op
		(ns)
		&
		Bandwidth
		(Gbps)
		\\
		\midrule
		
		\multirow{5}{*}{4}
		& DesignWare     & 0.98 & 378.52 & 375 & 367.50 & 1020 & 339.01 & 2.95 & 1.36 \\
		& Modified Booth & 1.21 & 460.71 & 206 & 249.26 & 826  & 274.11 & 3.65 & 1.10 \\
		& Dadda          & 1.23 & 442.89 & 203 & 249.69 & 813  & 269.65 & 3.71 & 1.08 \\
		& Wallace        & 1.05 & 455.13 & 237 & 248.85 & 952  & 315.88 & 3.17 & 1.26 \\
		& MRUD           & \textbf{0.85} & 462.57 & 298 & 253.30 & \textbf{1171} & \textbf{388.38} & \textbf{2.58} & \textbf{1.55} \\
		\midrule
		
		\multirow{5}{*}{8}
		& DesignWare     & 1.66 & 804.38 & 532 & 883.12 & 602 & 199.49 & 5.01 & 1.60 \\
		& Modified Booth & 1.75 & 858.65 & 590 & 1032.50 & 571 & 189.23 & 5.29 & 1.51 \\
		& Dadda          & 1.72 & 876.20 & 360 & 619.20 & 581 & 192.53 & 5.19 & 1.54 \\
		& Wallace        & 1.56 & 961.59 & 263 & 410.28 & 599 & 198.29 & 5.04 & 1.59 \\
		& MRUD           & \textbf{1.19} & 1080.49 & 365 & 434.35 & \textbf{840} & \textbf{278.72} & \textbf{3.59} & \textbf{2.23} \\
		\midrule
		
		\multirow{5}{*}{16}
		& DesignWare     & 2.24 & 2277.23 & 1114 & 2495.36 & 446 & 147.83 & 6.77 & 2.37 \\
		& Modified Booth & 2.37 & 2329.89 & 1347 & 3192.39 & 422 & 139.72 & 7.16 & 2.24 \\
		& Dadda          & 2.98 & 2332.55 & 893 & 2661.14 & 334 & 110.75 & 9.03 & 1.77 \\
		& Wallace        & 2.56 & 2542.96 & 1058 & 2708.48 & 400 & 132.46 & 7.55 & 2.12 \\
		& MRUD           & \textbf{1.68} & 2827.85 & 1758 & 2953.44 & \textbf{595} & \textbf{197.11} & \textbf{5.07} & \textbf{3.15} \\
		\midrule
		
		\multirow{5}{*}{32}
		& DesignWare     & 3.56 & 6514.34 & 3142 & 11185.52 & 281 & 93.02 & 10.75 & 2.98 \\
		& Modified Booth & 5.45 & 6031.28 & 1489 & 8115.05 & 183 & 60.76 & 16.46 & 1.94 \\
		& Dadda          & 5.53 & 7705.49 & 1305 & 7216.65 & 181 & 59.88 & 16.70 & 1.92 \\
		& Wallace        & 5.23 & 7991.17 & 1720 & 8995.60 & 182 & 60.21 & 16.61 & 1.93 \\
		& MRUD           & \textbf{2.14} & 9365.59 & 4925 & 10539.50 & \textbf{461} & \textbf{152.60} & \textbf{6.55} & \textbf{4.88} \\
		\bottomrule
	\end{tabular}
\end{table*}

\begin{figure}[pos=htbp]
	\centering
	\scriptsize
	\includegraphics[width=0.8\columnwidth, keepaspectratio=true]{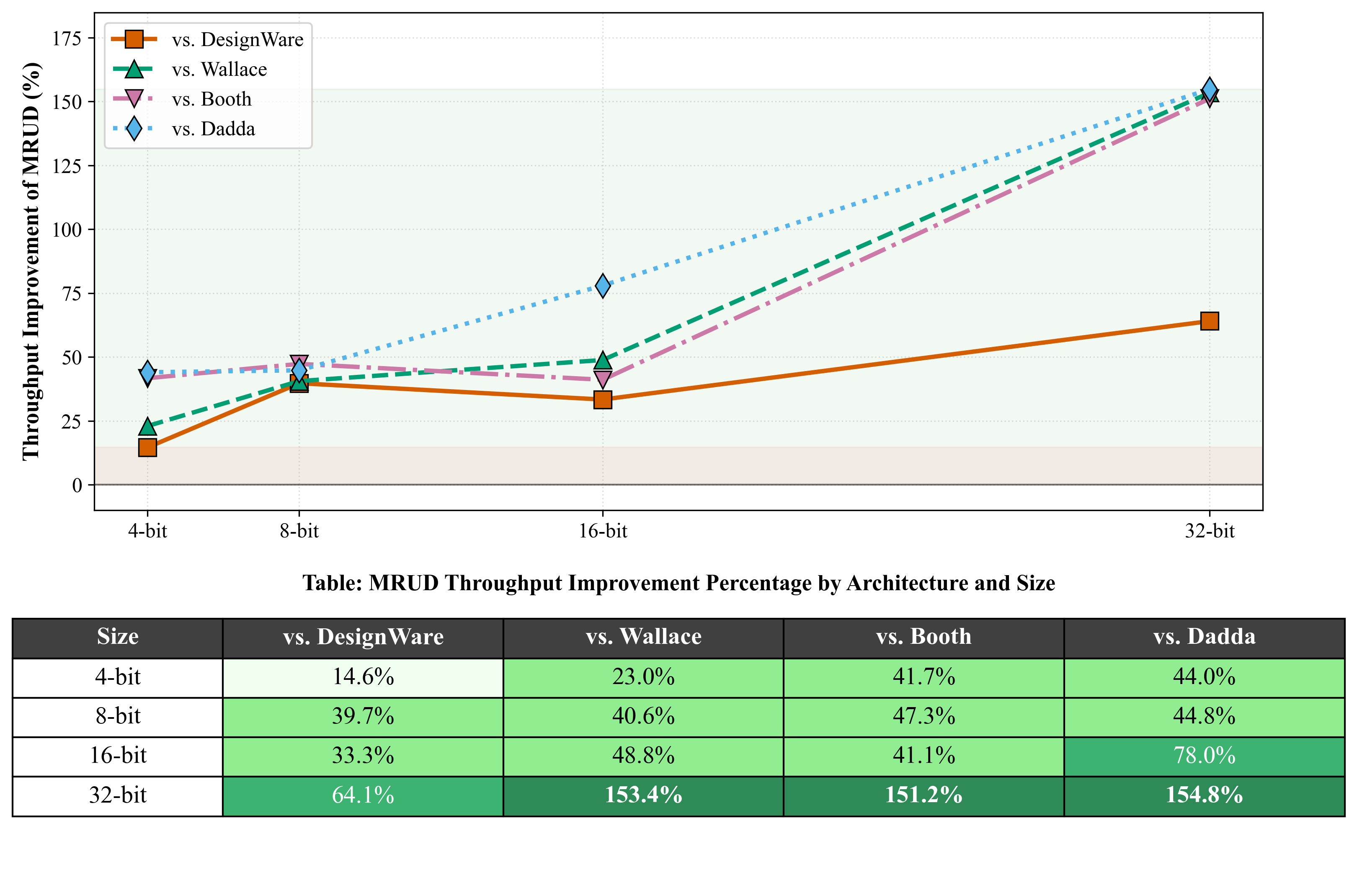}
	\caption{Throughput improvement of MRUD-based MAC over conventional architectures across different operand sizes.}
	\label{fig:throughput_improvement}
\end{figure}

To better visualize the performance advantage, Fig.~\ref{fig:throughput_improvement} presents the throughput improvement of the MRUD-based MAC compared to conventional architectures across all operand sizes. The improvement becomes increasingly pronounced for larger operand widths, reaching up to 64\% over DesignWare and approximately 154\% over Dadda for 32-bit operands, confirming that the benefit of eliminating carry propagation grows with operand size.

\begin{figure}[pos=htbp]
	\centering
	\scriptsize
	\includegraphics[width=0.8\columnwidth, keepaspectratio=true]{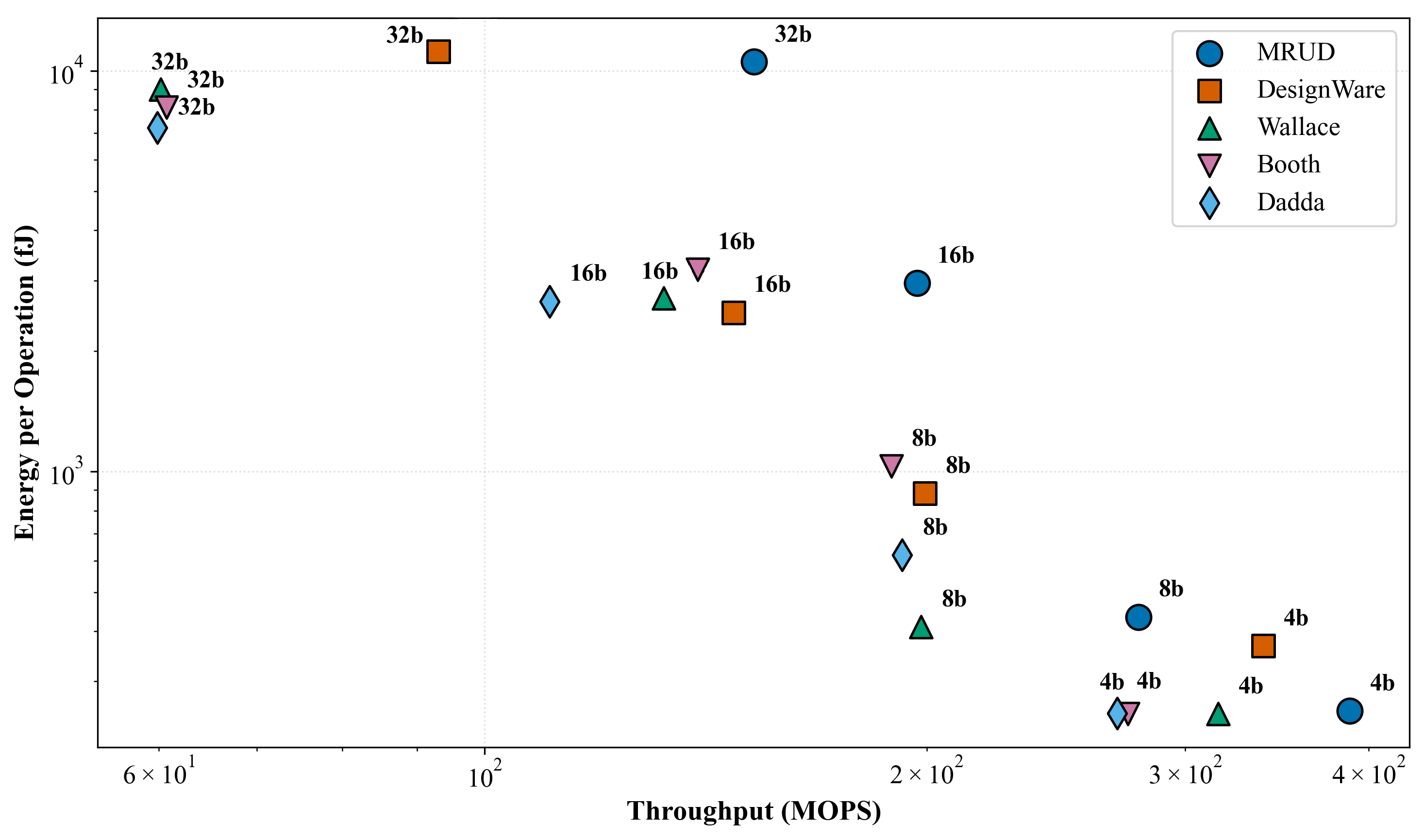}
	\caption{Energy per operation vs. throughput for different MAC architectures and operand sizes.}
	\label{fig:energy_throughput}
\end{figure}

The energy efficiency of the MRUD-based MAC is further analyzed in Fig.~\ref{fig:energy_throughput}, which plots the energy per operation against throughput for all evaluated architectures and operand sizes. The proposed MRUD architecture consistently achieves higher throughput while maintaining competitive energy efficiency compared to conventional designs. For 32-bit operands, the MRUD MAC delivers a throughput of 152.6 MOPS with an energy of 10,539.50 fJ, outperforming DesignWare (93.02 MOPS, 11,185.52 fJ) and Wallace (60.21 MOPS, 8,995.60 fJ). This confirms that the MRUD-based MAC offers a favorable energy-throughput trade-off across all operand sizes.

The area overhead of the MRUD MAC (9,365.59 µm² for 32-bit) is moderate compared to conventional designs (6,514.34 µm² for DesignWare), but is justified by the substantial performance gains. Although power consumption is higher (4,925 µW vs. 3,142 µW for DesignWare), the energy efficiency remains competitive due to the reduced execution time. The measured energy for 32-bit is 10,539.50 fJ, which is comparable to DesignWare (11,185.52 fJ) and Dadda (7,216.65 fJ), while offering significantly higher throughput.

\section{Conclusion}\label{sec:Conclusion}

This work introduced Maximally Redundant Unsigned Digit (MRUD) number systems for high-performance arithmetic circuits. By eliminating negabits, MRUD simplifies partial product generation and enables carry-free addition in higher radices, reducing logic complexity and critical path delay.

Key contributions include digit extension techniques for both signed and unsigned MRUD representations, a hardware-efficient sign inversion method with closed-form expressions, and a unified adder/subtractor architecture with comprehensive overflow/underflow detection that distinguishes real from apparent conditions. The proposed overflow/underflow detection scheme enables efficient result truncation by identifying when the most significant digit can be eliminated, reducing word length without compromising correctness.

A significant advantage of MRUD over MRSD is its inherent support for both signed and unsigned number representations. While MRSD is fundamentally restricted to signed arithmetic, MRUD, like conventional binary number systems, can be configured to operate in either signed or unsigned mode. This dual-mode capability makes MRUD particularly attractive for applications that predominantly involve unsigned operations, such as digital signal processing, image processing, and cryptography.

MRUD-based multipliers achieve up to 30\% lower delay, 26\% smaller area, and 87\% lower power than MRSD. In BBR mode, MRUD outperforms conventional architectures with up to 43\% delay reduction. The advantage is further amplified in MAC units, where reverse conversion is performed only once, yielding up to 64\% higher throughput. Although MRUD incurs moderate area overhead, the performance gains justify this trade-off. Future work includes exploring low-power MRUD implementations, applying the representation to division and square root operations, and investigating MRUD-based approximate computing for error-resilient applications.

\printcredits

\bibliographystyle{cas-model2-names}

\bibliography{refs}




\end{document}